\documentclass[12pt,preprint]{aastex}
\usepackage{graphicx}

\shorttitle{NIR Coronal Line in Nearby AGNs}
\shortauthors{Rodr\'{\i}guez-Ardila et al.}

\begin{document}


\title{The Near-Infrared Coronal Line Spectrum of 54 Nearby Active Galactic Nuclei}


\author{A. Rodr\'{\i}guez-Ardila}
\affil{Laborat\'orio Nacional de Astrof\'{\i}sica/MCT - Rua dos Estados Unidos 154, Bairro das Na\c{c}\~oes. CEP 37504-364, Itajub\'a, MG, Brazil}
\email{aardila@lna.br}

\author{M. A. Prieto}
\affil{Instituto de Astrof\'{\i}sica de Canarias (IAC), C/V\'{\i}a L\'actea, s/n, E-38205, La Laguna, Tenerife, Spain}
\email{aprieto@iac.es}

\author{J. G.  Portilla}
\affil{Observatorio Astron\'omico Nacional, Facultad de Ciencias, Sede Bogot\'a, Universidad Nacional de Colombia}
\email{jgportillab@unal.edu.co}

\and

\author{J. M. Tejeiro}
\affil{Observatorio Astron\'omico Nacional, Facultad de Ciencias, Sede Bogot\'a, Universidad Nacional de Colombia}
\email{jmtejeiross@unal.edu.co}



\begin{abstract}

The relationship between emission of coronal lines (CLs) and nuclear activity
in 36 Type~1 and 18 Type~2 active active galactic nuclei (AGNs) is analyzed,
for the first time, based on near infrared (0.8-2.4\,$\mu$m) spectra. The eight CLs
studied, of Si, S, Fe, Al and Ca elements and corresponding to ionization 
potentials (IP) in
the range 125 -- 450 eV, are detected (3$\sigma$) in 67\% (36 AGNs) of the sample.
Our analysis show that the four most frequent coronal lines − [Si\,{\sc vi}]\,1.963$\mu$m, 
[S\,{\sc viii}]\,0.9913$\mu$m,
[S\,{\sc ix}]\,1.252$\mu$m and [Si\,{\sc x}]\,1.430$\mu$m, − 
display a narrow range in luminosity, with
most lines located in the interval log$L$ 39 - 40 erg\,s$^{-1}$. We found that the 
non-detection is largely associated with either a lost of spatial resolution  
or increasing object distance: CLs are essentially nuclear and  
easily loose contrast in the continuum stellar light for nearby
sources or get diluted by the strong AGN continuum as the redshift    
increases. Yet, there are AGNs where the lack of coronal emission,  
i.e., lines with IP $\geq$ 100 eV, may be genuine. The absence of  
these lines reflect a non-standard AGN ionising continuum, namely, 
a very hard spectrum lacking photons below a few  
Kev.

The analysis of the line profiles points out to a trend of increasing
FWHM with increasing IP up to energies around 300\,eV, where a maximum in the
FWHM is reached. For higher IP lines, the FWHM remains nearly constant 
or decreases with increasing IP. We ascribe this effect to an increasing 
density environment as we approach to the innermost regions of these AGNs, 
where densities above the critical density of the CLs with IP larger than 
300\,eV are reached. This sets a strict range limit for the density in the 
boundary region between the narrow  
and the broad region of 10$^8 - 10^9$ cm$^{-3}$.

A relationship between the luminosity of the coronal lines and
that of the soft and hard X-ray emission and the 
soft X-ray photon index is observed: the coronal emission becomes 
stronger with both increasing x-ray emission (soft and hard) and steeper 
X-ray photon index, i.e.  
softer X-ray spectra.  Thus, photoionization appears as the dominant  
excitation mechanism. These trends hold when considering Type 1 sources only; 
they get weaker or vanish  when including Type~2 sources, 
very likely because the X-ray emission measured in the later is not the 
intrinsic ionising continuum.

\end{abstract}


\section{Introduction}

Coronal lines (CLs) are emission features arising  from forbidden transitions of  
excited states of highly ionized species  ($h\nu_{ion}\geq$ 100 eV), implying the 
existence of very energetic processes  occurring in or near the narrow line
region (NLR).  
Torus \citep{b18}, close or in part of the broad line region \citep{b64, b65},  
NLR \citep{b66, b67} and extended narrow line region (ENLR) 
\citep{b69, b68} have being proposed as possible locations for the formation of
these lines in active galactic nuclei (AGNs).  

CLs, also referred  by some authors as Forbidden High Ionization Lines (FHILs), are 
excited by collisions. What it is not clear  is 
the type of  mechanism  responsible of the high ionization of the species. Two basic
processes has been proposed in order to explain the high  level of ionization:   
photoionization due to the central source, which  emits  an intense  ionizing continuum,
particularly  hard UV and soft X-ray \citep{b3,b12,b4,b69,fer97,b5}; and shocks  between 
high velocity clouds   and the narrow line region gas, i.e.,  collisional ionization  
much alike of the process occurring in the solar corona  \citep{b6,b11}.  There exists 
a third intermediate possibility: a  combination of the above two processes \citep{b44}. 
Recent works conclude that although  nuclear photoionization  is the main excitation 
mechanism, in order to reproduce adequately the  observed emission line 
ratios,  a contribution of shocks is necessary \citep{b50,g09}. 

Since the mid 70s,  it has been claimed  that optical CLs tend to be broader than   
lower ionization  forbidden lines and its centroid position blueshifted
with respect to the systemic velocity of the galaxy  \citep{b12,b14,b4}.
Indeed, a correlation is  found between the ionization 
potential (IP) necessary to create  
the ionized specie and the line width
\citep{b15,b16,b70}. All together, this result is
interpreted as the coronal region be located between the narrow and the
broad line region and subjected to radial motions in some cases, probably
associated to outflows \citep{b4,b17}.

With the recent generation of high-sensitive detectors in the near infrared region (NIR),
observations of a significant number of Seyfert galaxies has now been possible 
\citep{b24,b28,b26,b27,b29,b30,g09,mul11}. As a consequence,
new CLs have been added up to list (i.e., [S\,{\sc viii}], [S\,{\sc ix}],
[Fe\,{\sc xiii}], [Si\,{\sc vi}], [Si\,{\sc vii}], [Si\,{\sc ix}], [Si\,{\sc x}],
[Ca\,{\sc viii}])  allowing a revision of previous findings and a more
consistent analysis of a complex emitting region. Among the most 
important results that can be mentioned are that the CLR is not restricted to the  
unresolved nucleus although it does not extended to the kpc range.  \citet{pri04,b50}, 
for instance, detect CLs  at distances of up to a few hundreds of parsecs from the 
nucleus. These authors also report that the highest ionized species  show  the 
smallest emitting region size of the NLR, which is in agreement with photoionization 
from the nucleus. These results gathered additional support by
the recent work of \citet{mul11}, who studied the spatial 
distribution and kinematics of the NLR and CLR of seven AGN usign SINFONI and OSIRIS near-IR 
AO-assisted integral field spectroscopy at spatial scales ranging from 4 to 36 pc. 
They found that the two regions present a bright compact core and extended emission,
with the CLR being more compact than the NLR, with typical sizes of the former between
80 and 150\,pc in radius. Moreover, both regions display a similar dispersion 
velocity although the high-ionization gas presents more deviant behavior with velocity 
fields dominated by non-circular motions, associated to outflows.

The aims of this paper are twofold. First, to carry out  the largest 
compilation of CLs  published to date in the near infrared region  (0.8-2.4 $\mu$m)
and second, from the study of this database, to draw clues about the excitation 
mechanisms of these lines by looking for a connection between their luminosity and 
broad-band measurements in other spectral regions. As the sample contains a 
significant number of Type 1 (Ty1, hereafter) and Type 2 (Ty2, hereafter) AGNs, 
a discussion of viewing angle 
effects and thus the most 
likely location of the CLR will be done.  This paper is structured as follows. 
Section~\ref{obs} describes the selection of the sample; Section~\ref{res} presents 
the main results, with emphasis of the relation between
FWHM and IP; the relation of CLs with soft and hard X-rays is 
discussed in  section~\ref{X-ray} and Section~\ref{con} contains the Summary and main
conclusions.

\section{Sample Selection}  \label{obs}

The  sample of  AGNs  chosen for this work was  selected primarily from the  galaxies  
studied  by \citet[RRAP06]{b30}.  It  includes  47 Seyferts and quasars
observed in the wavelength interval 0.8-2.4 $\mu$m and collected using cross-dispersed 
spectroscopy at IRTF. Description of the conditions of the observations, 
instrumentation  details  and  reduction process can be found in  RRAP06. 

RRAP06 sample was complemented with an additional sub-sample of active galaxies  
with published data on CLs covering a similar wavelength interval.
 These objects are: Circinus \citep{b23,  b101},  NGC 1068
\citep{b33, b100, b50}, Mrk 335, Mrk 1044, 
Ton S180, NGC 863 \citep{b27}, Mrk 766 \citep{b31}, Mrk 1210 \citep{b48} 
and Mrk 78 \citep{b51}. Basic information of the selected galaxies is shown 
in Table~\ref{tab1}. The objects 
are listed in increasing order of right  ascension. The final sample is composed of
54 AGNs: 36 of Ty1 (19 of which are narrow line Seyfert 1, NLS1) and 
18  of Ty2. Values  of redshifts 
were taken from the NED database. The last column indicates 
whether or not the galaxy displays at least one CL (at 3~$\sigma$ detection)
in the 0.8$-$2.4~$\mu$m region. Objects marked with  ``N''  do not show CLs 
in the the NIR spectrum. Yet, some of the them have CLs in other spectral ranges: 
these are: Mrk 334  
([Fe\,{\sc x}]~$\lambda 6374$, \citet{b94}  and 
[Ne\,{\sc v}]~14.3\,$\mu$m,  \citet{b28}),  NGC~1144 ([Ne\,{\sc v}]~14.3\,$\mu$m,  
\citet{b28}), NGC~2110 ([Fe\,{\sc vii}]~$\lambda \lambda 5721, 6087$  and 
[Fe\,{\sc x}]~$\lambda 6374$, \citet{b95}),   NGC~5929 ([Fe\,{\sc x}]~$\lambda 6374$, 
\citet{b94}) and Mrk~493 ([Fe\,{\sc vii}]~$\lambda 6087$,  \citet{b96}). 
All other AGN marked with ``N'', to our knowledge, do not display any coronal
lines either in the NIR or in any other wavelength region. 

\section{Results}  \label{res}

\subsection{Frequency of coronal lines}

\begin{figure*}     
\includegraphics[angle=90, width=13cm]{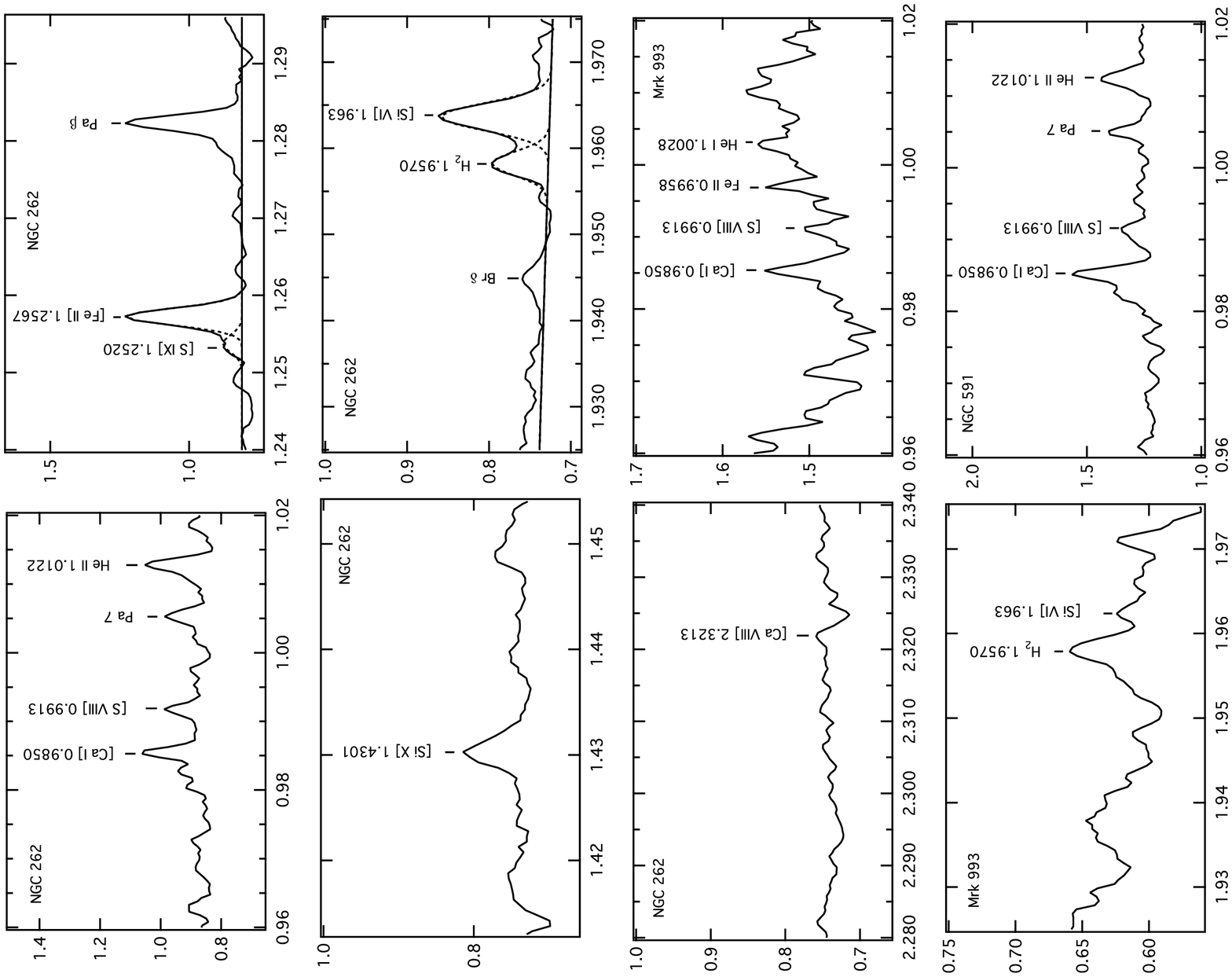}
\includegraphics[angle=90, width=13cm]{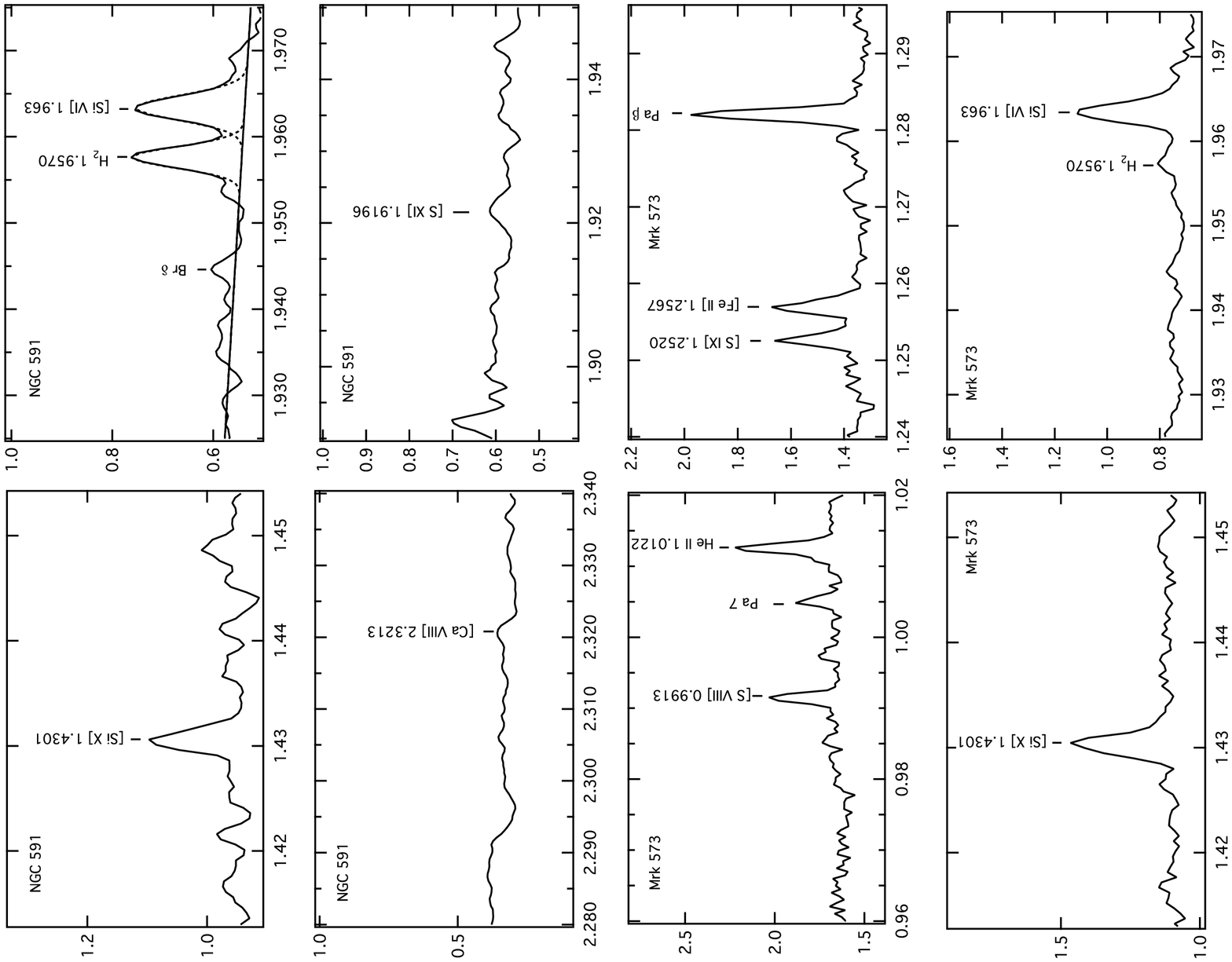}    
\caption{Zoom around the region containing the detected coronal lines in NGC~262, Mrk~993, NGC~591, and
Mrk~573. Wavelengths are in microns 
and fluxes in units of  $10^{-15}$ ergs\,cm$^{-2}$\,s$^{-1}$\,\AA$^{-1}$. For objects with 
strong blending emission lines, the Gaussian fits applied to separate the individual 
components are also shown. \label{fig-lay0}}   
\end{figure*}

\begin{figure*}     
\includegraphics[angle=90, width=13cm]{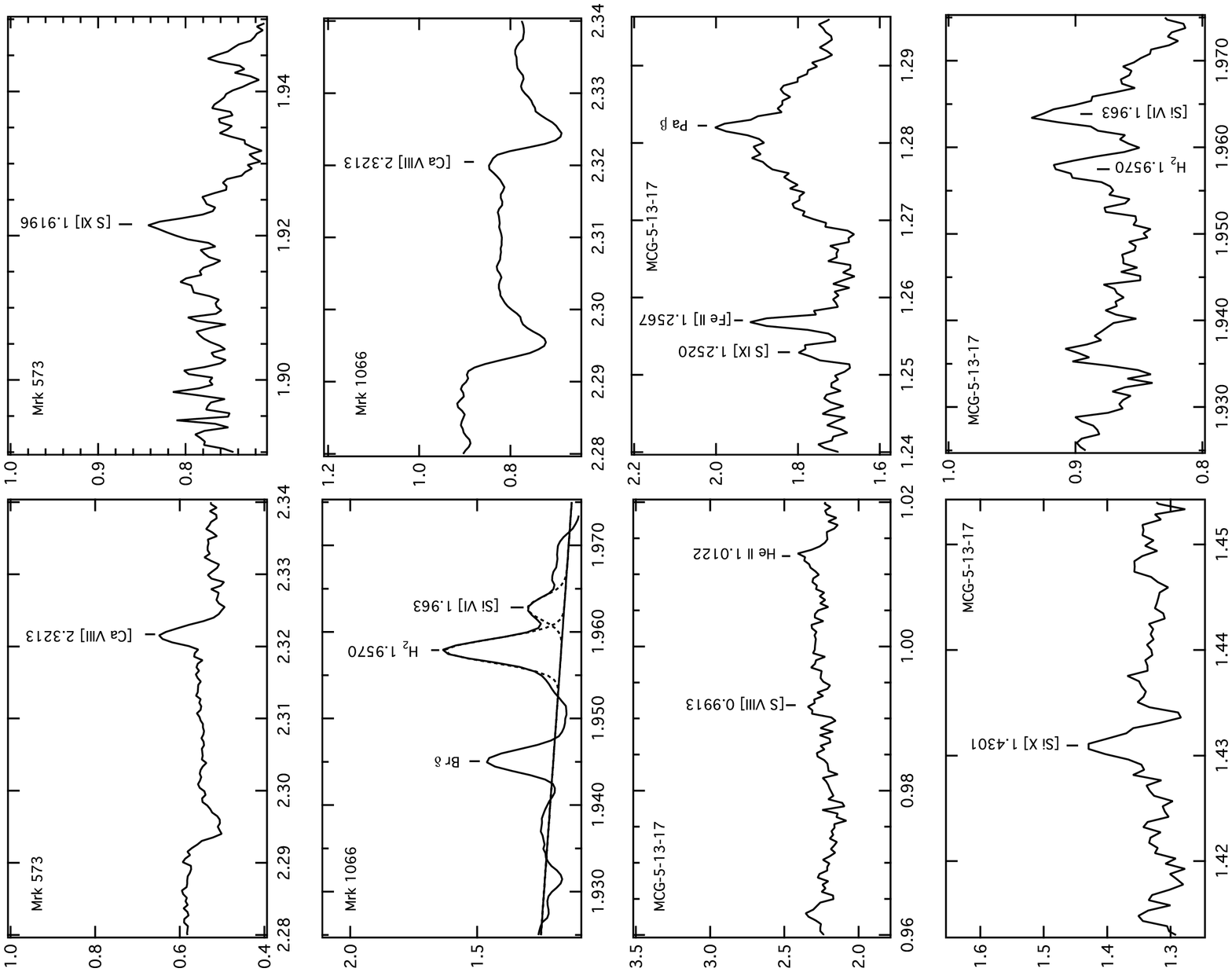}
\includegraphics[angle=90, width=13cm]{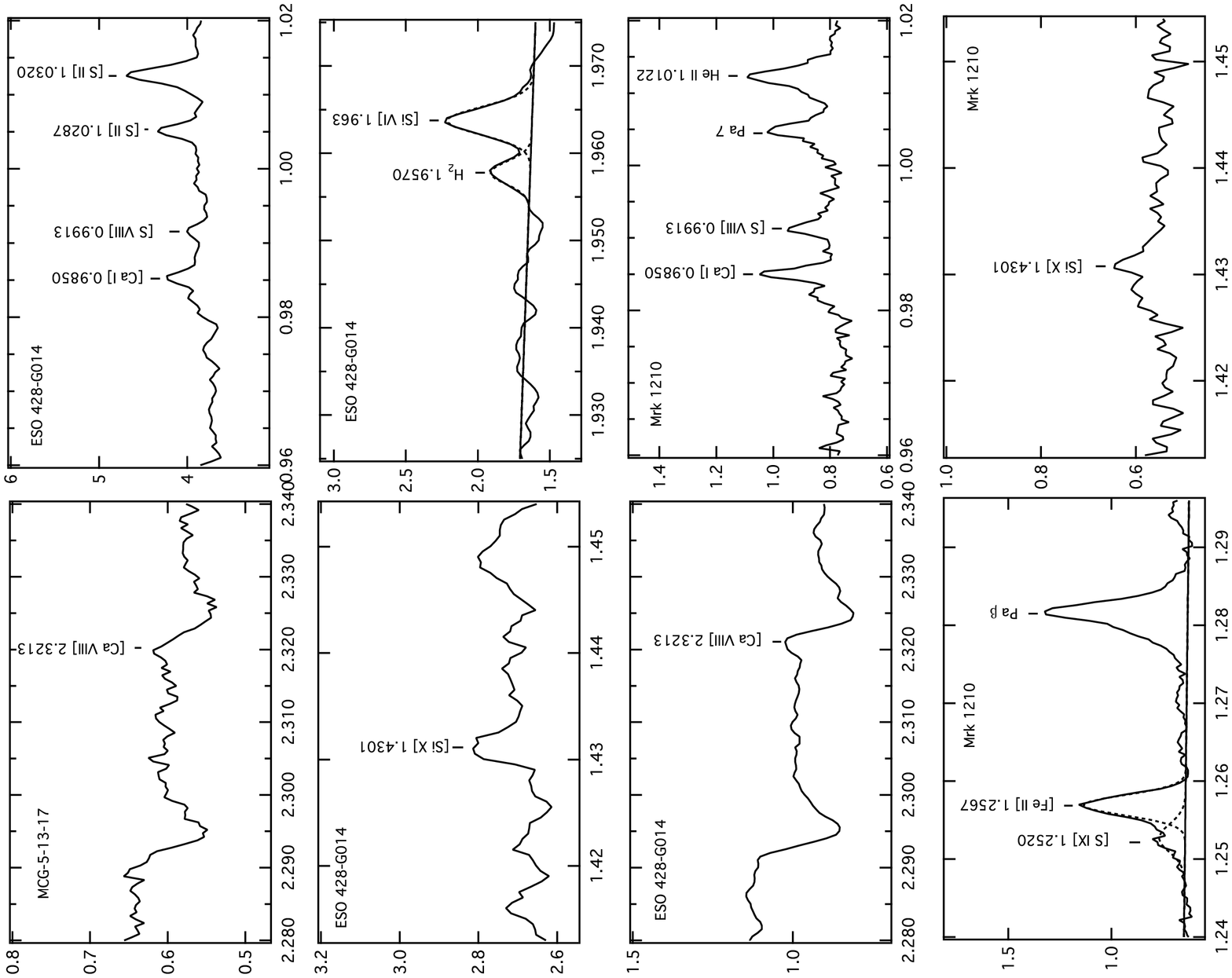}    
\caption{Same as Fig.~\ref{fig-lay0} for Mrk~573 (cont.), Mrk~1066, MCG-5-13-17, ESO~428-G014 and Mrk~1210. \label{fig-lay1}}   
\end{figure*}

\begin{figure*}     
\includegraphics[angle=90, width=13cm]{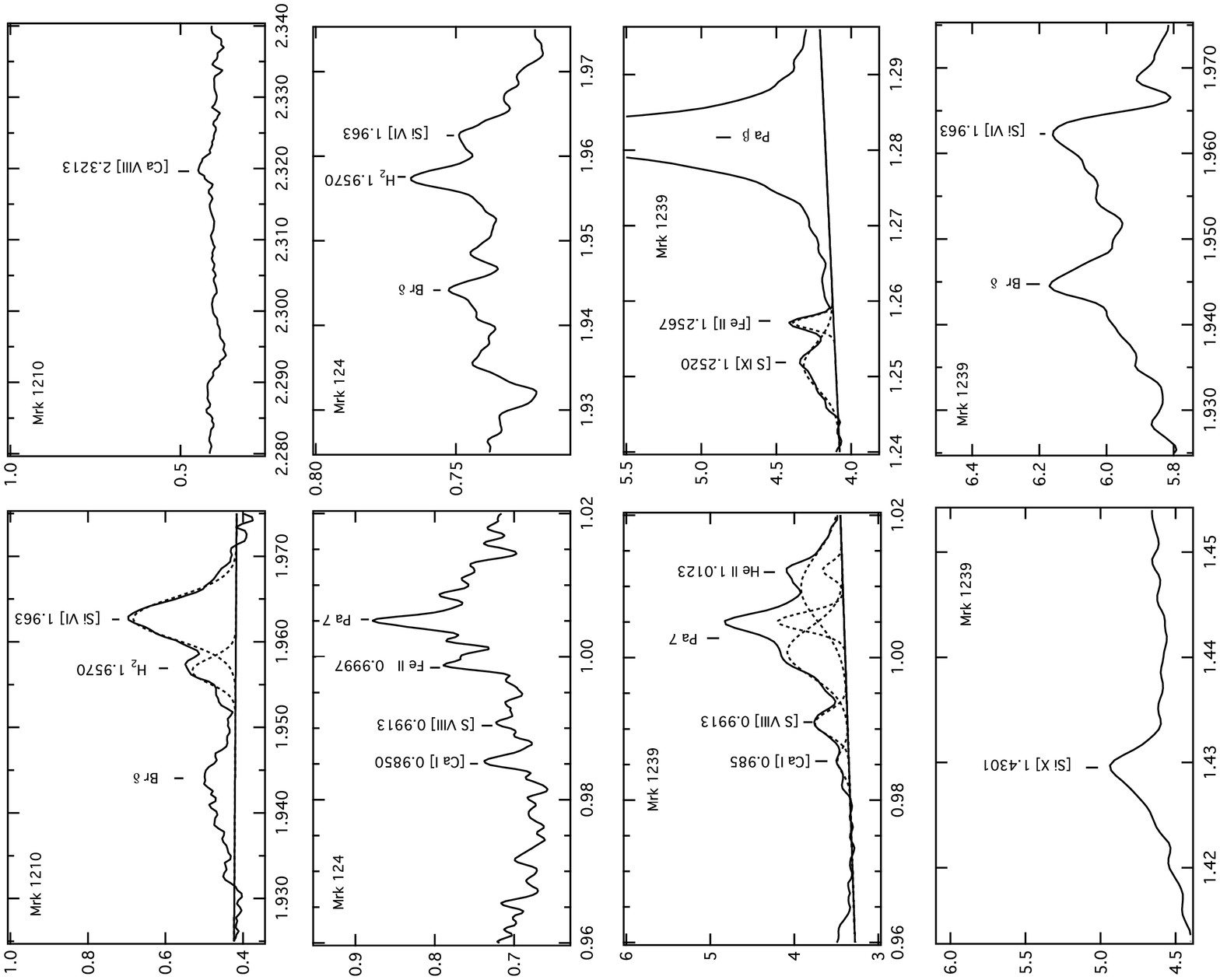}
\includegraphics[angle=90, width=13cm]{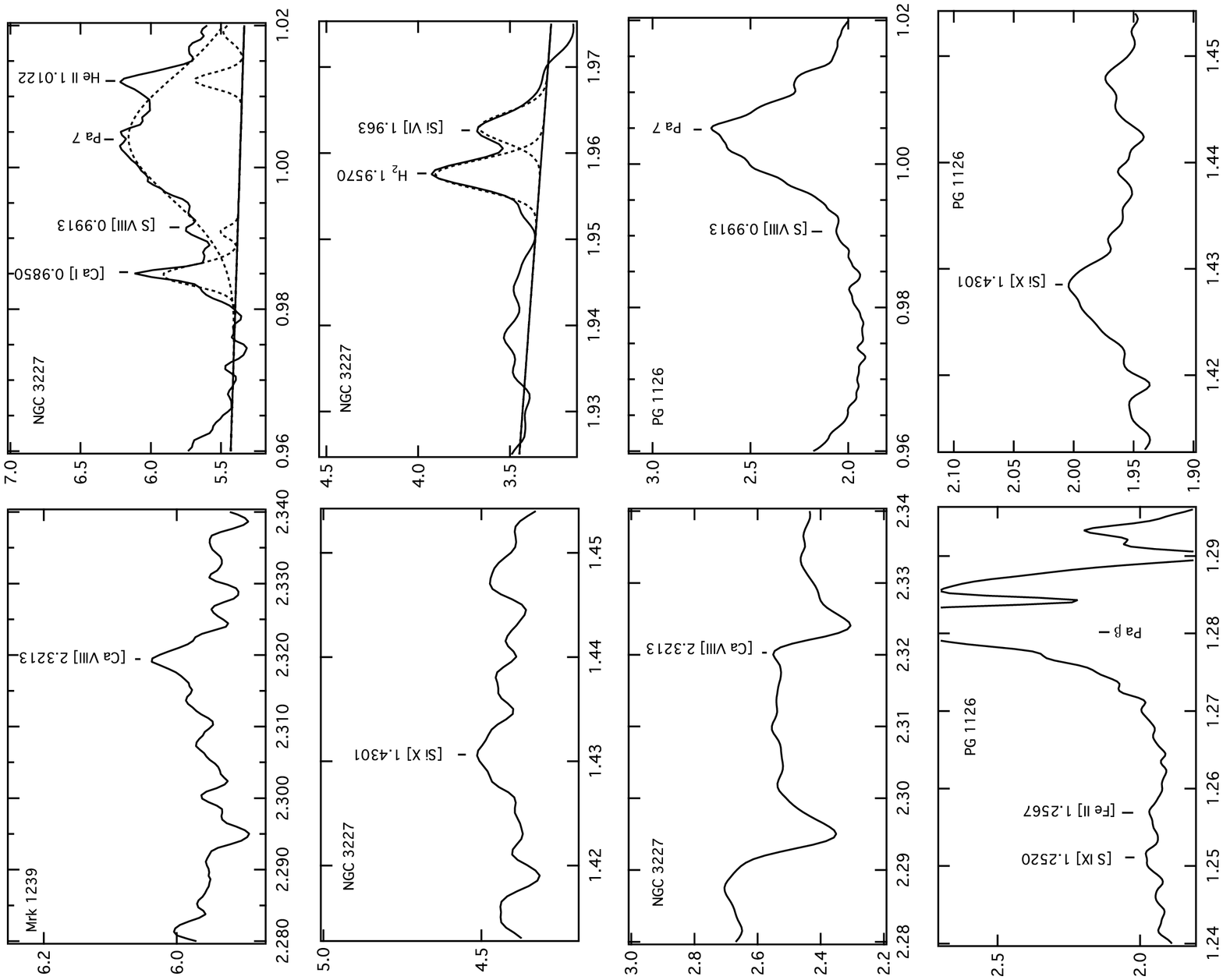}    
\caption{Same as Fig.~\ref{fig-lay0} for Mrk~1210 (cont.),   Mrk~124,  Mrk~1239, NGC~3227 and  PG1126-041  \label{fig-lay2}}   
\end{figure*}

\begin{figure*}     
\includegraphics[angle=90, width=13cm]{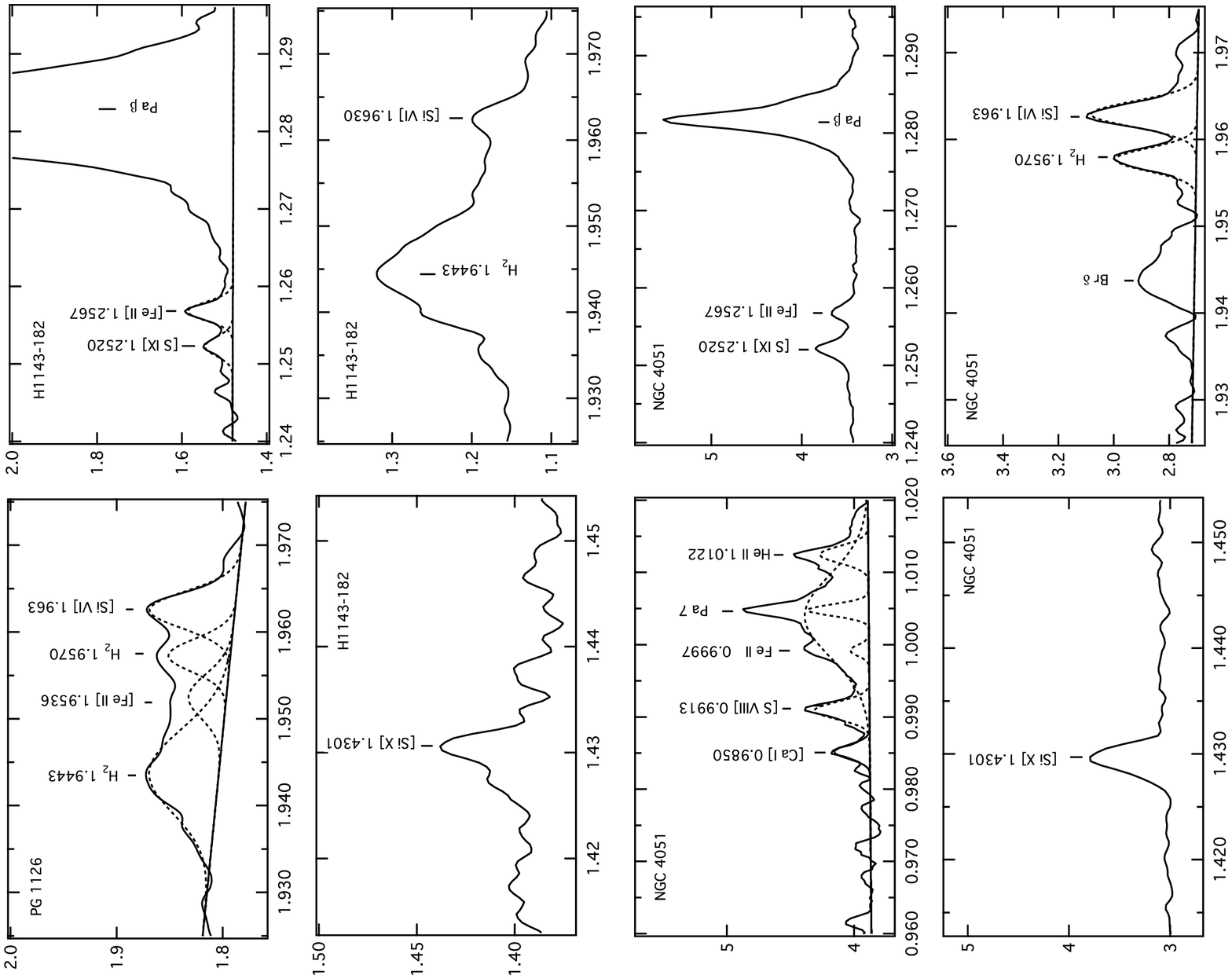}
\includegraphics[angle=90, width=13cm]{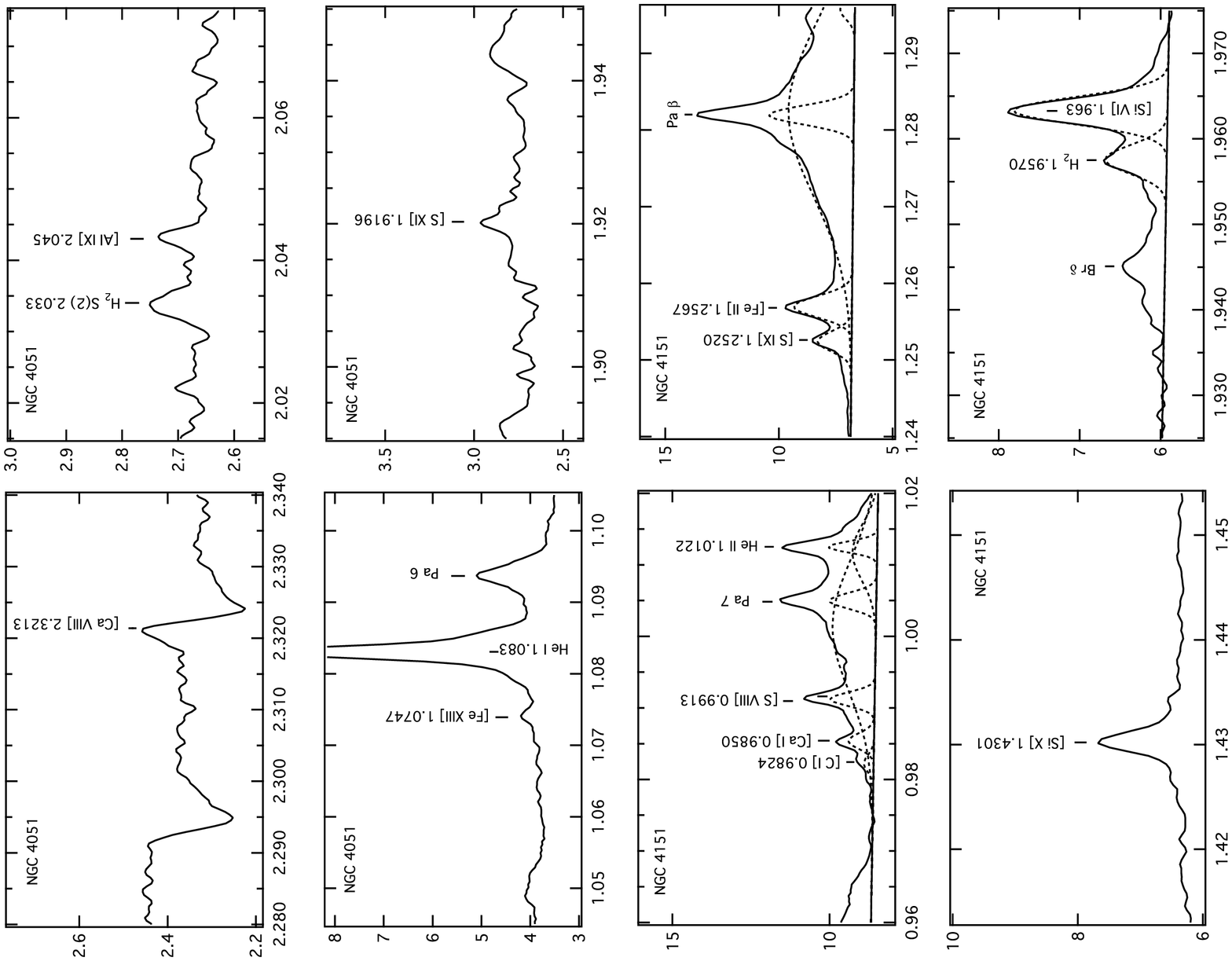}    
\caption{Same as Fig.~\ref{fig-lay0} for  PG~1126-041 (cont.), H~1143-182, NGC~4051 and NGC 4151.  \label{fig-lay3}}   
\end{figure*}

\begin{figure*}     
\includegraphics[angle=90, width=13cm]{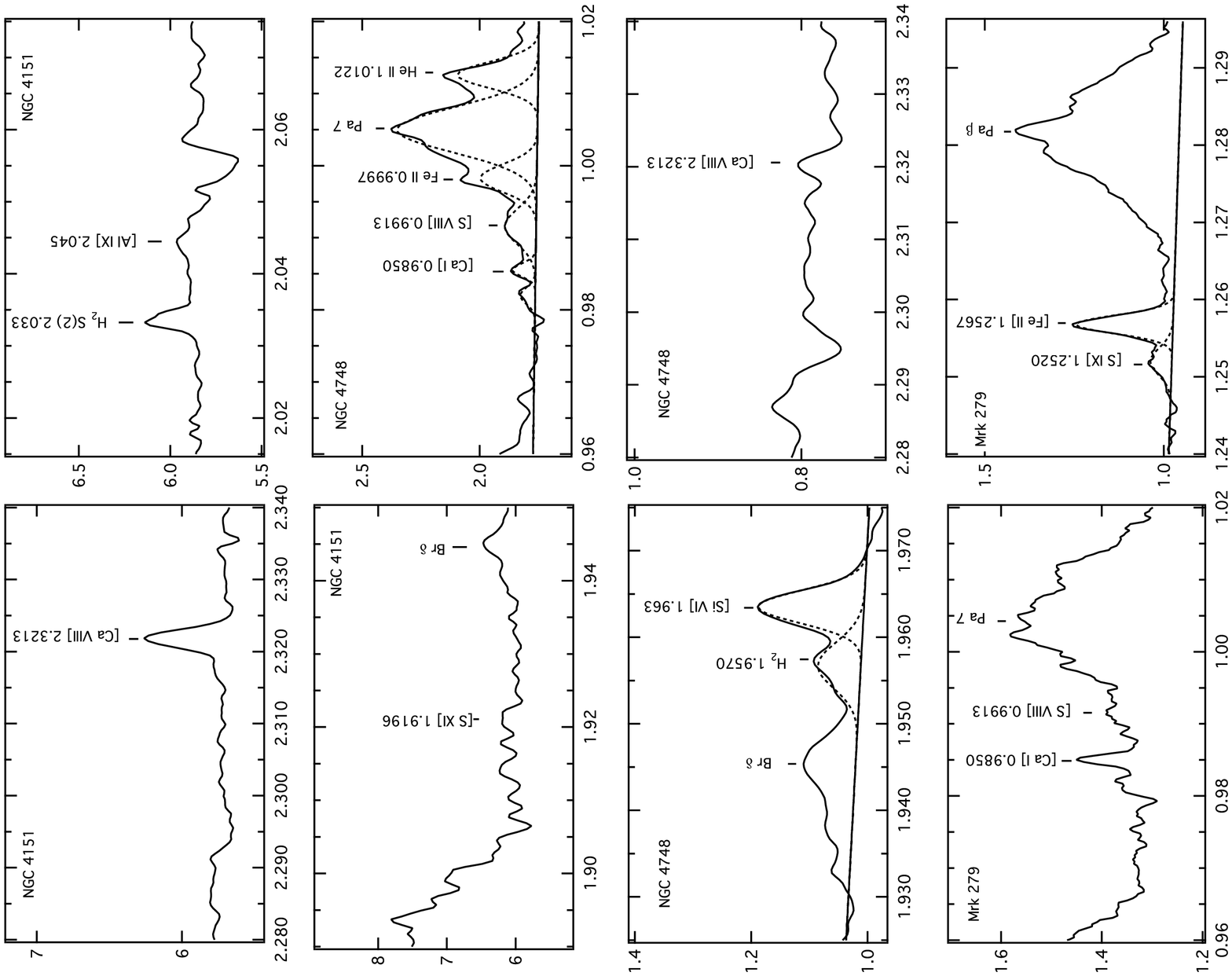}
\includegraphics[angle=90, width=13cm]{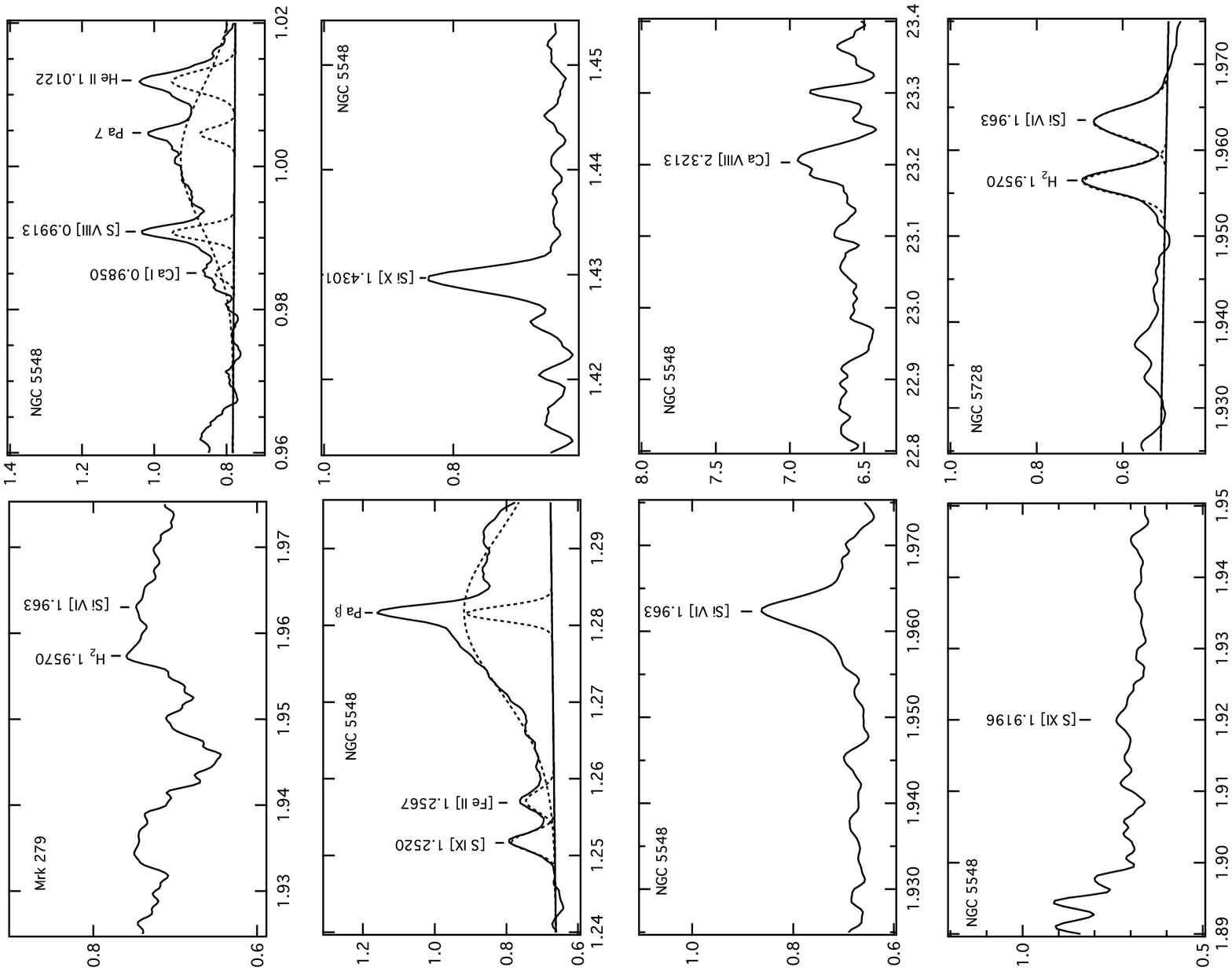}    
\caption{Same as Fig.~\ref{fig-lay0} for NGC~4151 (cont.), NGC~4748, Mrk~279,  NGC~5548 and NGC~5728.  \label{fig-lay4}}   
\end{figure*}

\begin{figure*}     
\includegraphics[angle=90, width=13cm]{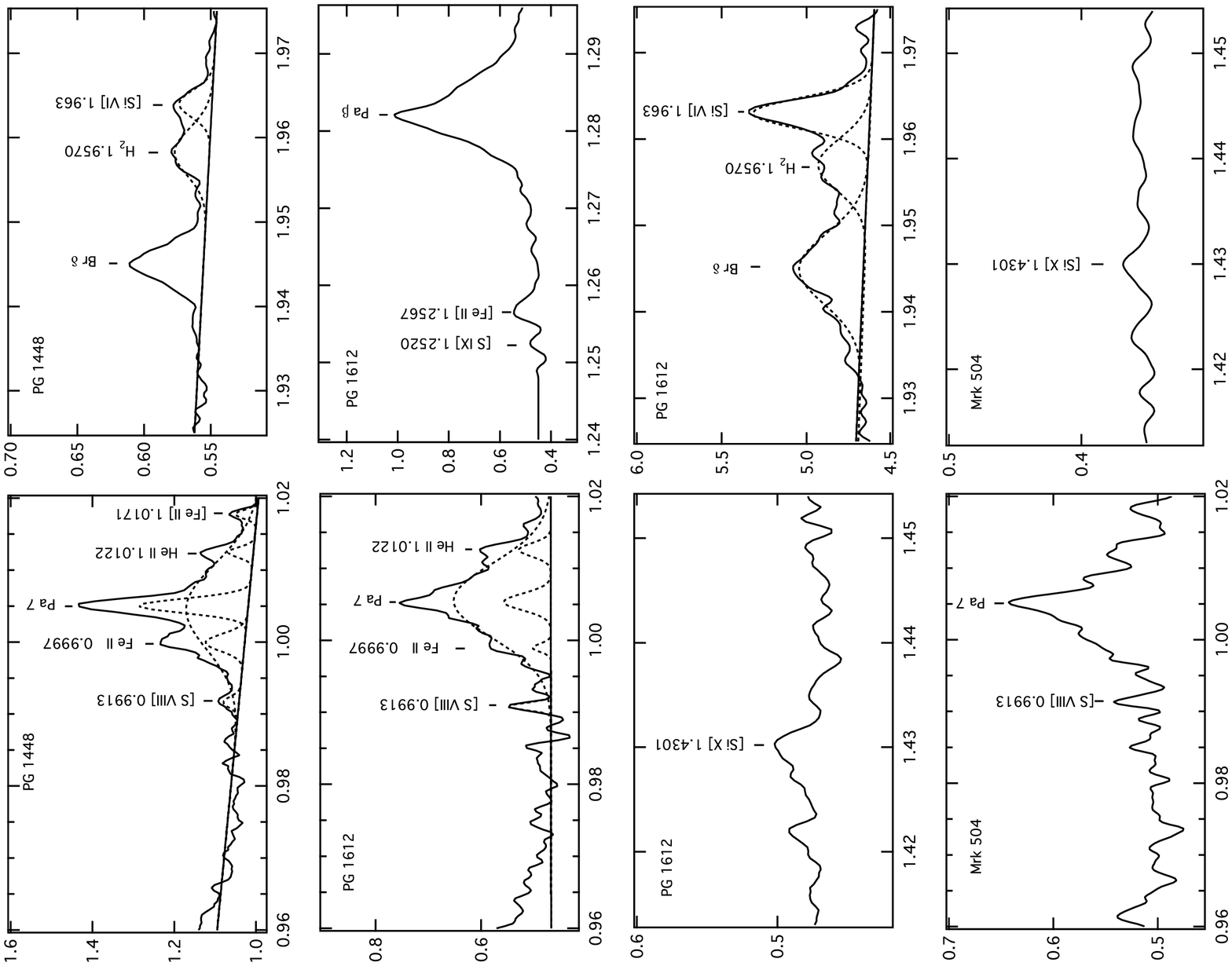}
\includegraphics[angle=90, width=13cm]{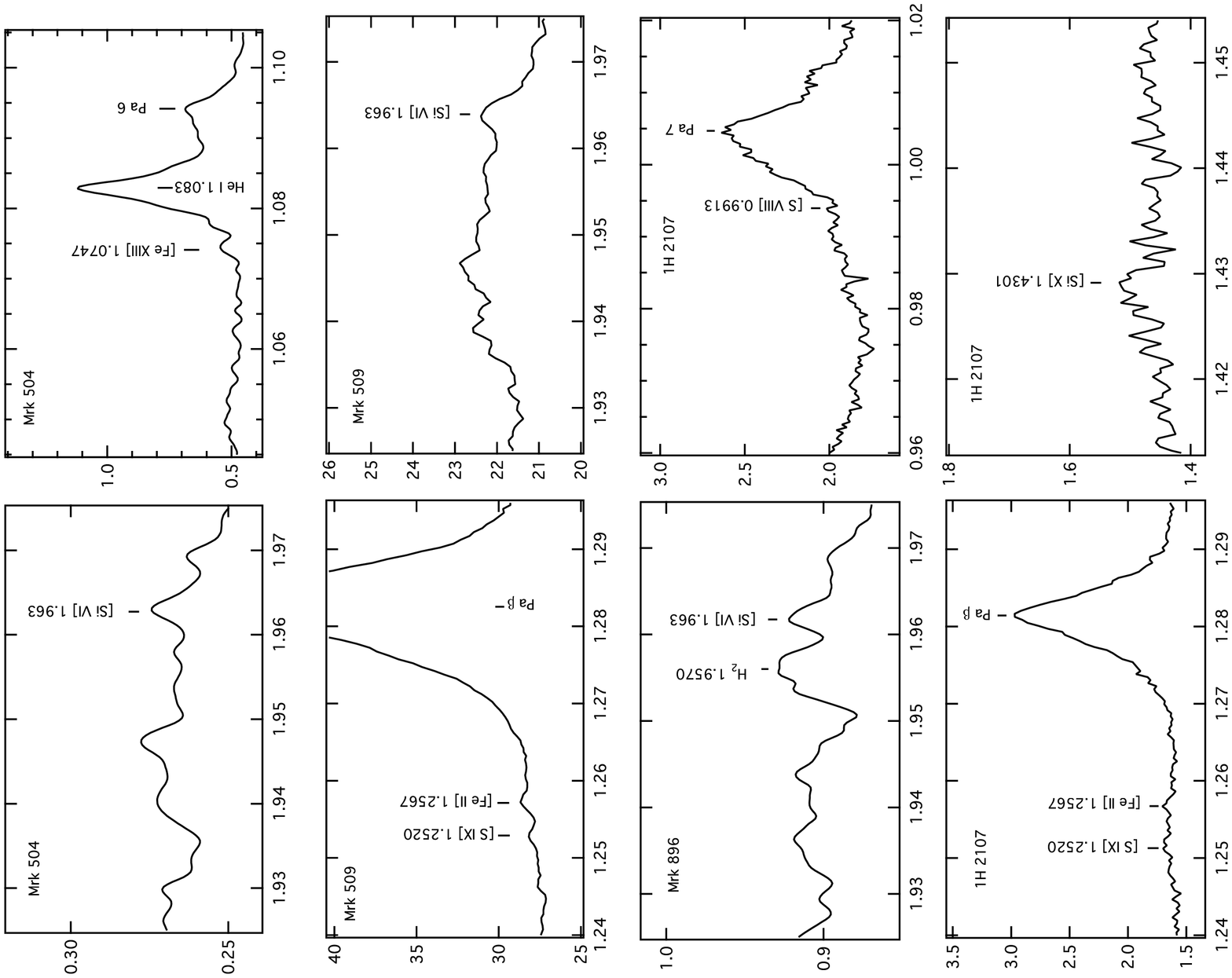}    
\caption{Same as Fig.~\ref{fig-lay0} for PG~1448+273, PG~1612+261,  Mrk~504, Mrk~509, Mrk~896 and 1H 2107-097.  \label{fig-lay5}}   
\end{figure*}

\begin{figure*}     
\includegraphics[angle=90, width=12cm]{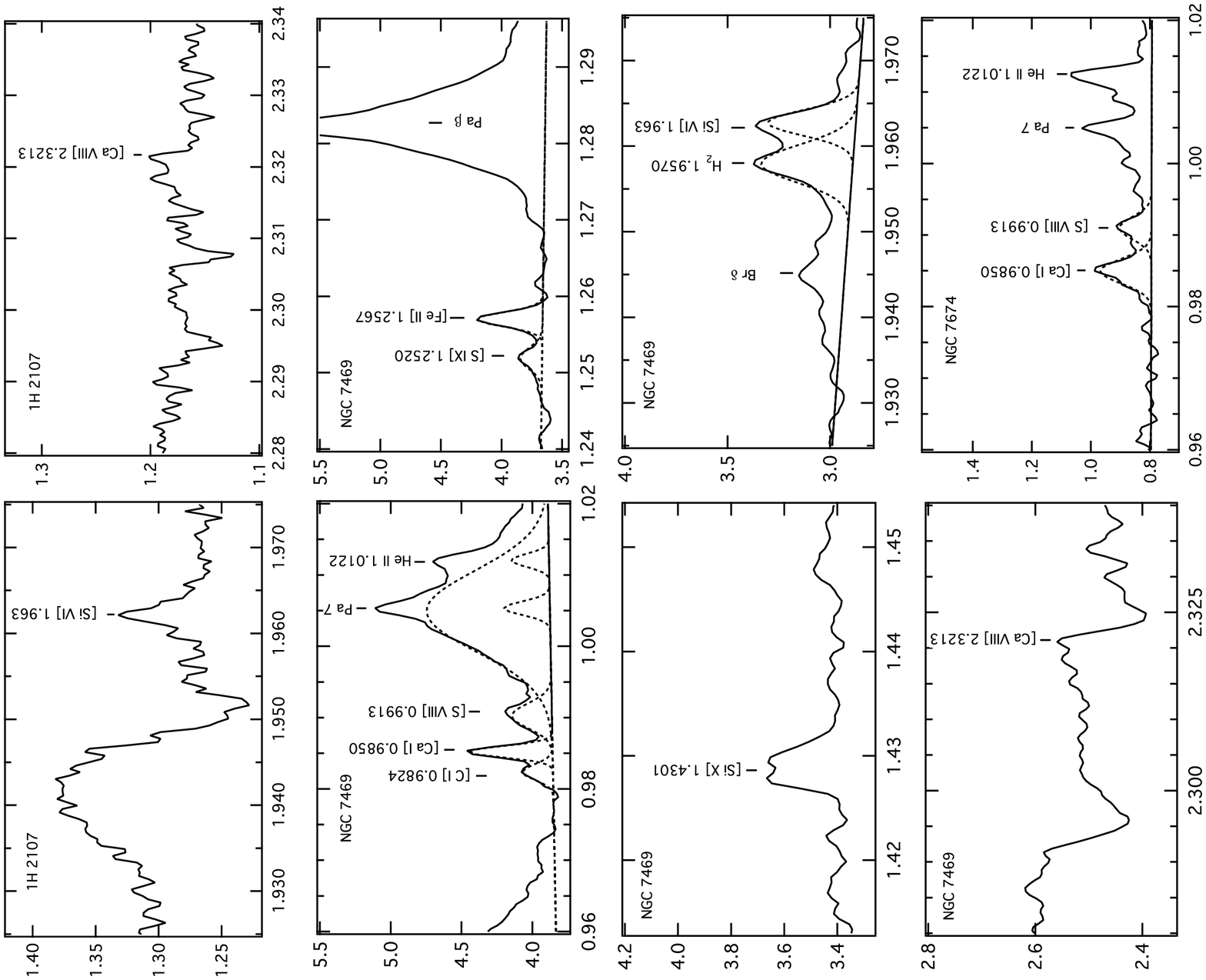}
\includegraphics[angle=90, width=10cm]{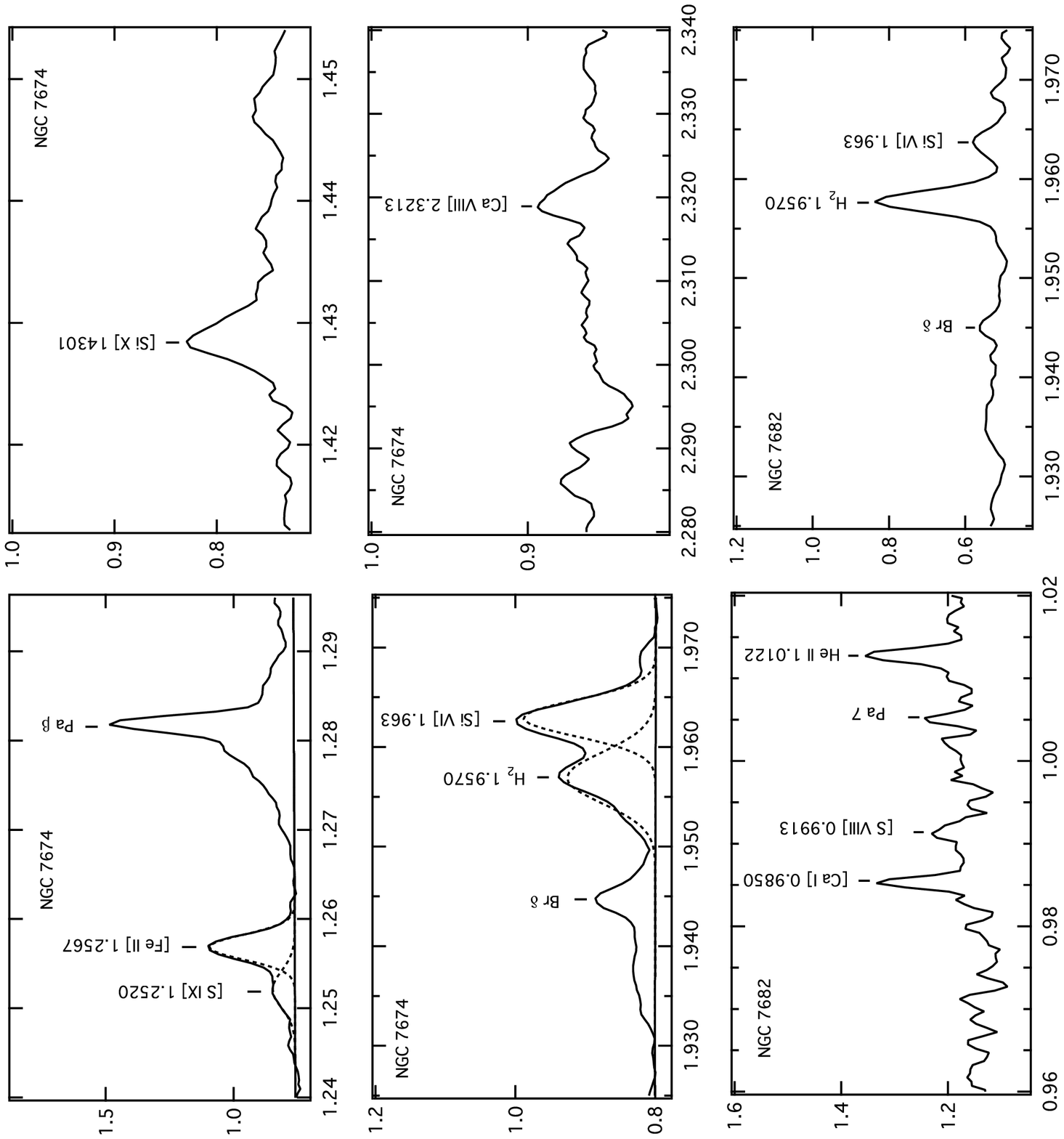}    
\caption{Same as Fig.~\ref{fig-lay0} for 1H 2107-097 (cont.),  NGC~7469, NGC~7674 and NGC~7682.  \label{fig-lay6}}   
\end{figure*}

The CLs that fall in the spectral window 0.8$-$2.4 $\mu$m are 
[S\,{\sc viii}]~0.9913~$\mu$m,  [Fe\,{\sc xiii}] 1.0747 $\mu$m,  
[S\,{\sc ix}]~1.252~$\mu$m,   
[Si\,{\sc x}]~1.4301~$\mu$m, [S\,{\sc xi}]~1.9196~$\mu$m, 
[Si\,{\sc vi}]~1.9630~$\mu$m,  [Al\,{\sc ix}]~2.0450~$\mu$m and  
[Ca\,{\sc viii}]~2.3213~$\mu$m (See Tab~\ref{tab2}). Few objects in the sample
display all them simultaneously. For each object with at least one CL detected,
Figures~\ref{fig-lay0} to~\ref{fig-lay6} show a zoom of the spectral region 
containing the CL. We found that 36 out of 54 galaxies (66\%) display at least
one CL. Of the remaining 18 sources, five of them show CLs in other 
spectral ranges (optical and/or mid-IR), as reported in the literature
(see Sect.~\ref{obs}). Adding these 5 objects, the number of AGN with at 
least one CL - i.e.  from ions with IP $\geq$ 100 eV - in any range of the 
spectrum increases to 41, that is, 76\% of the  sample.

Table~\ref{tab70} lists the CLs fluxes measured in sample along with fluxes
of the narrow components of Pa$\alpha$ and Br$\gamma$, except when
indicated. Because of discrepancies
either in the line fluxes or in the uncertainties relative to the values published by RRAP06
in some cases, we have reviewed all their measurements and these are the ones presented
in Table~\ref{tab70}. In cases of line blending, a multiple Gaussian
fit was applied. Upper limits are reported for the expected position of the lines 
if they were not detected above 3$\sigma$ level. 

It is easy to see from Table~\ref{tab70} that for all 3-$\sigma$ detections, 
the most frequent and conspicuous CLs in the sample are those of silicon 
and sulfur, and among those, the ones with the lowest IP ($<$ 300 eV) are 
the strongest. CLs from other elements, e.g., Ca, Al or Fe, are less
frequent and weaker, the former two being affected by metallicity effects 
and depletion onto dust, while the later may be not be that much affected. 
Indeed, the iron optical coronal lines are rather  strong. [Fe\,{\sc vii}]\,6087\AA\ 
for instance, is even stronger than the strongest NIR CL [Si\,{\sc vi}]\,1.963$\mu$m 
(e.g. \citet{b50}). Yet,  the [Fe\,{\sc xiii}]\,1.0747$\mu$m line traced in 
this work is located in wing of the strongest NIR line, 
He\,{\sc i}\,1.0830$\mu$m, which most probably hampers its detection in 
most cases. Overall, [Si\,{\sc vi}]\,1.963$\mu$m, [S\,{\sc viii}]\,0.991$\mu$m, 
[Si\,{\sc x}]\,1.431$\mu$m and [S\,{\sc ix}]\,1.252$\mu$m are the lines 
with the largest number of detections, being observed in 30, 24, 22 and 
20 objects, respectively. For that reason, the rest of the analysis 
shall focus on these lines only.

\begin{figure}    
\plotone{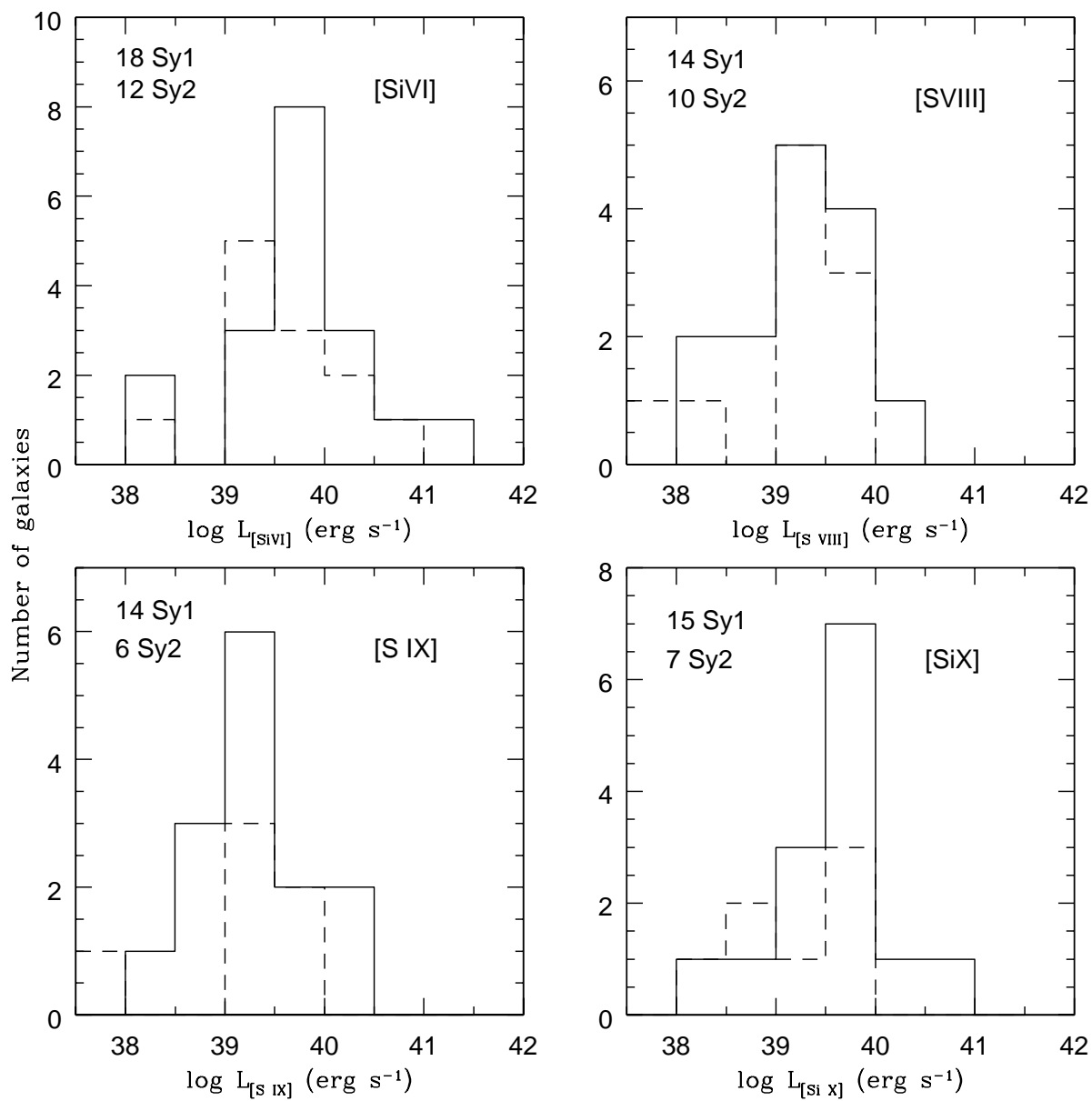}   
\vspace{0.1cm}
\caption{Histogram with the number of galaxies for each bin of
log in luminosity (full line for Ty1 and dashed line for Ty2 AGNs) for the 
four most frequent coronal lines found in the sample. The total numbers of Ty1 and Ty2
AGNs used for every line are indicated in the upper left corner of
each panel.   \label{figm}}    
\end{figure}

Figure~\ref{figm} shows a histogram with the number of galaxies per bin in
luminosity (full line for Ty1 and dashed line for Ty2) in each of the above 
four lines. Upper limits are not included. Broadly speaking, 
the distribution in luminosity for both types of AGNs is similar, 
with most lines located in the interval log$L$ 39-40 erg\,s$^{-1}$
and the whole sample spanning 3~dex in full-width around that interval. 
Note that the numbers above are similar to the range of 
[Fe\,{\sc x}]~6374~\AA\ luminosity reported by \citet{gel09} using a sample 
of 63 AGN with strong high-ionization emission lines, none of them common
to our sample. Figure~\ref{figm} also suggests a similar percentage
between the two types of AGNs that display [Si\,{\sc x}] and [S\,{\sc ix}]
($\sim$40\% and $\sim$35\%, respectively). Ty2, however, slightly outnumbers 
Ty1s in the lower ionization CLs 
[Si\,{\sc vi}] and [S\,{\sc viii}]: 67\% vs 50\% and 56\% vs 39\%, respectively.
These differences can be attributed to the relatively higher redshift - $z$ - 
of Ty1 objects, 
making the detection of CLs more difficult due to the poorer spatial resolution 
sampling. In addition, when $z > 0.035$, both lines
start falling in low atmospheric transmission zones or strong telluric absorption,
hindering their detection.
Note that only one Ty2 of the sample (Mrk~78) has $z>0.03$ 
while $\sim$35\% of Ty1s exceeds this redshift.

Similar results on the frequency of coronal line   
emission in Seyfert galaxies regardless of their type have been  
reported in the literature, albeit on the basis of smaller samples.    
Most of these works are conducted in the NIR \citep[e.g][]{b28,b29,b50,b52} 
and mid -IR \citep[e.g.][]{b28}. They all sample CLs from elements 
including Si and Mg, which could get  easily depleted from the gas phase 
if dust is present. Yet, all these works show that Ty2 objects show as rich and  
strong CL spectra as those of Ty1.
However,  analysis  done on the basis of the optical CL spectrum,  
mainly on iron lines, indicate opposite results. Works done  
by \citet{b41,b40} and \citet{gel09}, for instance, remark a significant 
weaker optical CL in Ty2
than in Ty1s. These authors argue that the frequency and  
strength of the CL spectrum  should thus depend on the orientation of  
the torus with the line of sight, being in Ty2 sources partially  
or totally obscured. As a consequence, the CLR should form at the  
inner face of the torus. 

Two relevant issues may explain these conflicting results. 
First, the extinction in the NIR is  
a factor 10 lower than in the optical. As Ty2 sources  
are more absorbed, this has a definitive impact on the detection and  
the measurement of optical CLs, mostly considering that they are 
relatively weak features. Indeed, a recent discovery of a Ty2 quasar
with an outstanding high-ionization spectrum and
zero internal extinction \citep{ros11} supports this argument. 
In moving to the NIR, the situation changes dramatically for strongly
reddened sources.   
Prototype Ty2 AGNs such as NGC\,1068 or Circinus, not only show the  
full spectrum of CLs (i.e., Si, Al, Ca, among others), but in addition,   
these are the dominant lines in their spectra \citep{b93,b29,mul09,mul11}.  

Second, spatial resolution should be taken into account in  
Ty2s for CL detection, especially in the optical but also in the  
NIR. Circumnuclear star formation, which is often more relevant in Ty2s,  
increases the continuum level with the corresponding  lost of  
line contrast. A stoning illustration of this effect is the nucleus  
of Centaurus A. It shows no CLs either in optical or in the NIR in seeing  
limited conditions \citep{b29}. But observations with  
adaptive optics reveal the K-band CLs of [Si\,{\sc vi}]\,1.963\,$\mu$m and 
[Ca\,{\sc viii}]\,2.34\,$\mu$m \citep{neu07}.  

Additional arguments against the CLR   
being obscured by the torus can be gathered from Circinus,  
NGC\,1068 and Centaurus~A. They all are extreme Compton thick sources even in the  
hard  X-rays. Under that condition, it is difficult that any optical CL 
emission could  be seen if their bulk is produced in the inner face of the 
torus. Moreover, the high spatial resolution studies conducted by \citet{b50} and 
\citet{m10} show that Ty2 sources 
display strong [Fe\,{\sc vii}], [Fe\,{\sc x}] and 
[Fe\,{\sc xi}] lines, with emitting regions extending to distances 
of a few hundred parsecs from the centre in the case of NGC\,1068.
That is, well beyond the
expected size of the putative torus. Moreover, a dusty CLR
implies in a very weak coronal line spectrum because of the absorption of the
incident continuum by dust at large U(H) \citet{fer97}. 
Although we do not rule this later hypothesis for those objects 
where no coronal lines are seen, at least those that displays CLs
should have the bulk of the CLR outside the torus.

\begin{figure}     
\plotone{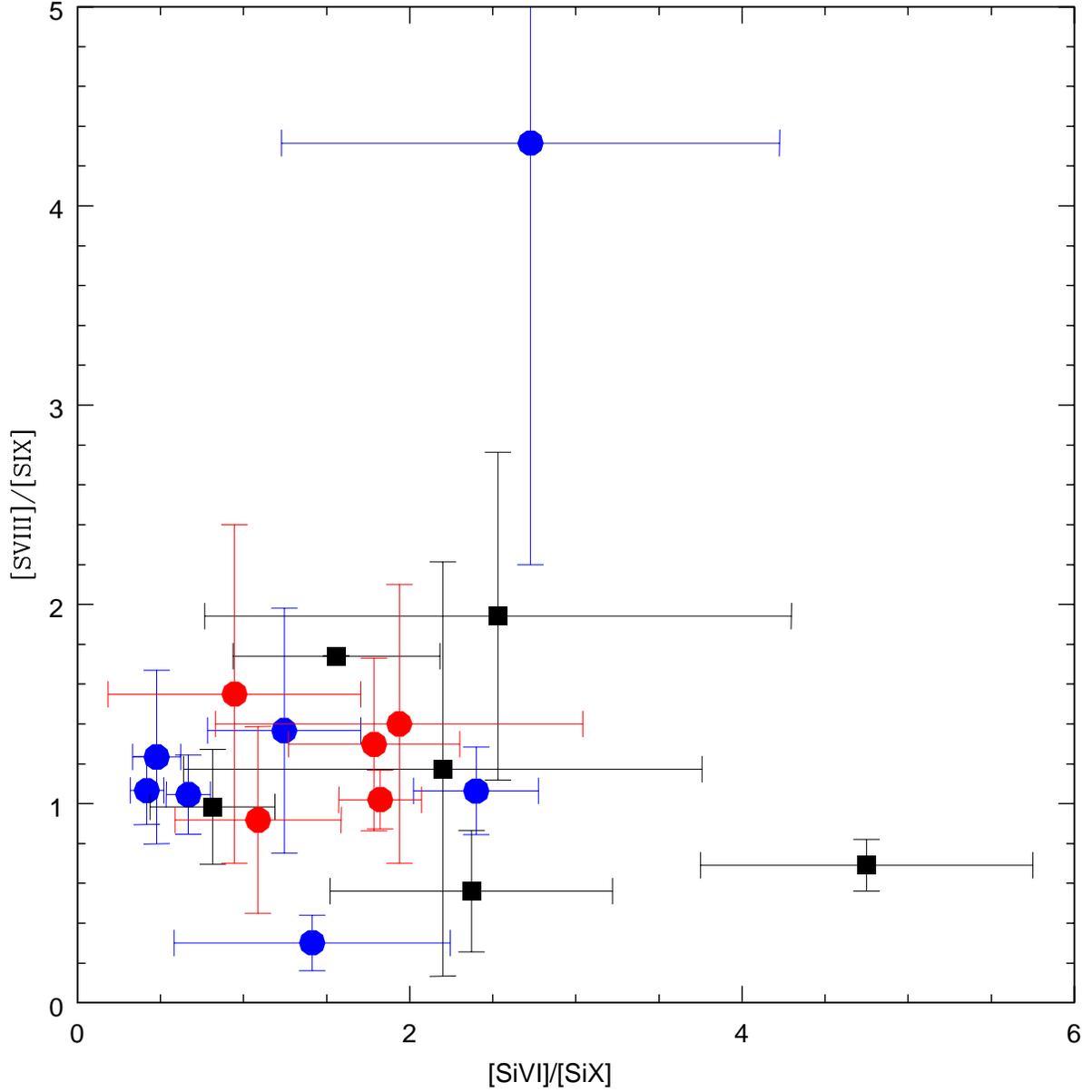}   
\vspace{0.1cm}
\caption{Flux ratio [Si\,{\sc vi}]/[Si\,{\sc x}]
versus [S\,{\sc viii}]/[S\,{\sc ix}] measured for the objects where at
least 3 of the fluxes involved was detected at 3$\sigma$ level. Circles
correspond to Ty1s (classical and NLS1) and squares to Ty2 AGNs. 
\label{ratio} }   
\end{figure}

In spite of the difference in coronal line luminosity between the galaxies, 
covering 3~dex in each of the four lines considered (see Fig.~\ref{figm}), the flux ratios 
[Si\,{\sc vi}]/[Si\,{\sc x}] and [S\,{\sc viii}]/[S\,{\sc ix}]
are distributed in a very narrow interval of values, as can be seen in
Fig~\ref{ratio}. Moreover, no distinction between Ty1 and Ty2 AGNs is
observed in both ratios, giving additional support to our previous finding of 
similar frequency of CLs in both type of objects. Note that these two ratios are independent of 
the metallicity as they contain ions of the same element. The 
abscissa is strongly sensitive to the form of the ionizing continuum because the 
difference in IP between [Si\,{\sc vi}] and [Si\,{\sc x}] is 184~eV 
(see Table~\ref{tab2}) while the ordinate ([S\,{\sc viii}]/[S\,{\sc ix}])
is not: the difference in IP between both ions is only $\sim$50~eV. 
It is easy to see that the scatter is the former 
ratio is restricted to 0.7~dex, with an average value of the ratio equal to 1.4. 
For the later, it is no larger than 0.4 dex, with an average of 1.1. 
Clearly, the bulk of CL fluxes  should be 
produced outside the torus region because the two ratios and their 
scatter are very similar in Ty1 and Ty2 AGNs.  
Observational evidence of this scenario was recently obtained 
by \citet{mul11}. They quantified the extent of the CLR by means of azimuthal averages 
within circular radial annuli of the flux distribution 
of [Si\,{\sc vi}] in a sample of seven AGNs, four of them common to
our sample (Circinus, NGC\,1068, NGC\,4151 and NGC\,7469). Their results
show that the flux distribution is resolved and extended, supporting an origin
of the CLs in the inner part of the NLR or in the transition region between the 
BLR and the NLR.

Recent studies \citep{whi05,schl09,m10} have pointed out 
photoionization as the main driver for 
the coronal lines. Yet, we note that detailed  
analysis  of CL flux ratios as done in \citet{b50} on  
the basis of [Fe\,{\sc x}]/[Fe\,{\sc vii}] vs [Fe\,{\sc xi}]/[Fe\,{\sc vii}]
or [Fe\,{\sc xi}]/[Fe\,{\sc x}] show that  
photoionization alone is not enough. Photoionization models predict  
the general trend of these ratios but specific  values are not  
matched even if considering new calculations of collision strengths  
for iron. \citet{b50} show that adding the 
extra power of shocks of different velocities to photoionization 
improves the match between the observed iron ratios and those predicted.
Note that AO observations of the CLR carried out by \citet{mul11} 
support this picture. They found kinematic signatures of outflows
in NIR high-ionization lines of a sample of nearby AGNs, with the radio jet 
clearly interacting with the ISM, indicative of a link between jet power 
and outflow power.

The lack of coronal lines above certain energies 
may be related to a deficient ionizing continuum.
This would explain why some objects display [Si\,{\sc vi}] (IP=166 eV) and
[S\,{\sc viii}] (IP=280 eV) but not lines of higher IP.
Examples of this among the nearest objects in the sample are  Mrk~1066,   
NGC~5728 - both lacking lines with IP $>$ 166 eV or  NGC~3227 and NGC~7682 - both  
lacking lines of IP $>$ 280eV (see Table ~\ref{tab70}).
A similar argument applies to some of the objects studied by  
\citet{m10} on HST/STIS spectra:
some of their objects, including NGC~3227, display [Ne\,{\sc v}] and [Fe\,{\sc vii}], 
both with IP $\sim$ 100~eV, but not
higher optical ionization lines. The results for NGC~3227 are further supported  by 
\citet{b50}, who report  
on the non detection of  [Fe\,{\sc xi}]~7892\AA\ ( IP=260 eV).

Distance should also play a major role in the detection of CLs: these are  
relatively weak features and loose of  
contrast over the stellar continuum light in low-luminosity nearby AGNs
hampers their detection. This argument is supported by results derived
from \citet{rif09}, who modeled the NIR continuum emission of
AGNs in terms of stellar population, 
AGN featureless continuum and hot dust in 9 Ty1 and 15 Ty2 galaxies. These 24 sources
are also studied in this work using the same set of data.
Cross-crossing Riffel et al.'s results with ours allow us to conclude
that for most Ty1 and Ty2 objects where
no CLs are detected, the contribution of the stellar population dominates the observed
NIR continuum \citep[see Table 3 of][]{rif09}. For example, in the Ty1 sources
NGC\,1097 and Arp\,102\,B, 96\% and 89\%, respectively, of the observed
continuum is due to stars. Similarly, in the Ty2s NGC\,1144, NGC\,5929 and 
NGC\,5953, more than 90\% of the observed continuum is due to that component.

In high-luminosity AGNs, the strong continuum emitted by the
central source would also dilute the CLs,
in a similar way the stellar population affects the CLs in nearby sources. Note 
that the effect of dilution, in this case, gets severe 
with increasing $z$. Observational evidence of this hypothesis can
be gathered by examining the work of \citet{gli06}, who constructed
the first quasar composite spectrum in the NIR using a sample of 27 
quasars with $z$ between 0.118 - 0.418. A close examination to 
Glikman et al.'s data shows that no coronal lines are observed
either in the individual sources 
nor in the composite template. Indeed, the only forbidden 
line that is clearly detected is [S\,{\sc iii}] 0.906~$\mu$m.
Under the assumption that photoionization by the central engine is
the dominant source of gas excitation, dilution by the AGN continuum
would preferentially affect lines emitted by the NLR.
Thus, the line flux ratio Pa$\alpha$/[S\,{\sc iii}] 0.906~$\mu$m
can be used as a metric for this effect. Note that the
total flux (narrow + broad) of Pa$\alpha$ can be employed here as 
the broad component of that line dominates the line flux in Ty1 
sources.

The value of the above ratio measured in the composite NIR quasar 
spectrum of \citet{gli06} is 12.4 (See Table 6 of that work). 
Dilution would tend to increase that value.
In our sample, nearby Ty1 sources have ratios approaching 4 at 
most (1H\,1934-063, NGC\,4748 for instance) while PG\,1612+261 
($z$=0.1309) has a ratio of 11.6. In very luminous objects 
like Mrk509 ($z$=0.034), that ratio reaches to 30. These results
evidence that the AGN continuum indeed dilutes the features emitted
by the NLR in luminous sources. Note that [S\,{\sc iii}] 0.906~$\mu$m 
is the strongest forbidden line observed in the NIR, being at least 5$\times$ 
stronger than [Si\,{\sc vi}]~1.963~$\mu$m, the strongest coronal
line of our survey.
  
The $z$ range of our sample is 0.0015 $<~z~<$0.5, median $\sim$0.02, 
1$\arcsec \sim$420 pc (Table~\ref{tab1}), and in general, most objects with 
$z~ < $0.005, 1$\arcsec ~<$ 110 pc, show   
CLs, although not always in the full range of ionization potentials.  
Above this $z$, there are a size able number of objects with coronal  
emission, some with a  remarkable spectrum, e.g. Mrk\,573 (Ty2), 1$\arcsec \sim$360\,pc,  
whereas others at equivalent or inferior distances do not show  
emission despite of the high Signal-to-Noise ratio (S/N) of the spectra, eg. NGC\,1275 (Ty2) or 
NGC\,1144 (Ty2), both at similar distance as Mrk\,573 (Table 2). 

All the above suggest that aside physical parameters (ionizing continuum,
metalicity, gas density or dust) observational issues like S/N or the spatial 
scale in the sky, strongly influence the detection of coronal lines. 
In the following sections we will collect information from the
NIR to increase our knowledge of the
physics of the CLs to complement the picture already
gathered from the visible region. 

\subsection{Kinematics of the coronal line region}

This section explores the  correlation between the ionised  gas  
velocity, as measured by the line FWHM, and the  ionization level of  
the gas, as measured by the IP of the corresponding line. If  
photoionization is the major mechanism, a correlation between these  
two parameters is expected. This is because  the highest the gas  
ionization level (high IP), the closer to the nucleus  is its  
location, therefore, the greatest is its motion within the gravitational
potential wells of the super massive black hole and host galaxy.  
Coronal lines are very suitable to study this prediction given the  
high range of IPs they cover. Most of the results published to date 
have been drawn from the study of 
small number of objects and CLs. The larger number of galaxies examined 
here and the inclusion of eight  CLs of different ions and wide range of 
IP allow us to investigate the FWHM-IP relationship in more detail.

In the late 70s,  \citet{b15}  reported  for NGC~3783 a  strong correlation between 
FWHM and ionization potential of various optical forbidden lines, with the CLs 
showing the largest FWHM and low or medium ionization lines the smallest ones.
Later works reported a similar trend in other active galaxies with either optical 
or NIR lines \citep{b36, b16, b4, b35}. 
Conversely, \citet{b24} found  
that CLs show  an ample variety of FWHMs, leading to the conclusion that the CLR occupies 
different regions in different galaxies. \citet{b34,b29} study the CLs in the 
1.5-2.5~$\mu$m range in 14 Seyfert galaxies. In only four 
of them a positive correlation between FWHM and IP could be established.
\citet{b27} in a sample 
of six galaxies (5 classified as  NLS1), found a positive correlation in three of them. 
Other observations have also shown that  
CLs can display no correlation at all or even an anti-correlation between FWHM
and IP. Such is the case of NGC~4151 
\citep{b37},  where [S\,{\sc xi}]~1.9196~$\mu$m  has a FWHM significantly lower 
than the narrow component of Pa$\beta$. Similarly, \citet{b33} reported for  NGC~1068
FWHM values for [Fe\,{\sc xi}]~0.7892~$\mu$m  and   [Si\,{\sc ix}]~3.9350~$\mu$m 
smaller than those of [Fe\,{\sc ii}]  1.257~$\mu$m whereas \citet{b50} and this work 
(Table~\ref{tablex2}) find the opposite: low- and intermediate- ionization lines 
present FWHMs  smaller or at least comparable to  those of the   optical   and   
near-IR  CLs.

Table~\ref{tablex2} lists the FWHM of the CLs (corrected from instrumental 
width, 360\,km~s$^{-1}$) along with the FWHM 
of [S\,{\sc iii}]~0.953 $\mu$m and [Fe\,{\sc  ii}]~1.257 $\mu$m,
two of the strongest medium- to low-ionization lines detected in the galaxy sample.   
In order to measure the FWHM the emission lines were assumed to be well 
represented by a single or a combination of Gaussians profiles 
(see Figs.~\ref{fig-lay0} to ~\ref{fig-lay6}). The LINER routine 
\citep{b39} was used in this process. 

Figures \ref{figpi} and \ref{figu} show plots of FWHM vs IP 
for each of the galaxies in the sample 
that display at least two coronal lines. For comparison purposes, the FWHM
of the lower ionization lines [S\,{\sc iii}]~0.953$\mu$m and [Fe\,{\sc ii}]~
1.257$\mu$m are also included. It can be  
seen that in some of the AGNs with sufficient number of CLs, a trend 
of increasing FWHM with IP is observed but up to energies of 
about 200$-$300 eV. In following this trend, we focus on the strongest CLs, namely those
from  Si and S, and ignore those of Fe and Ca as they are overall weaker and  
their FWHM more difficult to measure or subjected to larger uncertainties.
At higher energies the FWHM, in particular of the most  
conspicuous and frequent lines, namely [S\,{\sc ix}] and [Si\,{\sc x}],
remains constant or start decreasing with 
increasing IP. These are, for example, the cases of Mrk~335, ESO~428-G14, 
Mrk~1210, Mrk~1239, NGC~4051, Mrk~766, Ark~564, NGC~7469, NGC~7674 and 
tentatively, Circinus,  We have labeled  these objects as  ``P''  
(peak at about 300 eV) in the last column of Table \ref{tablex2}.  
In contrast, NGC~4151 and NGC~5548, two of the galaxies with the highest S/N 
in the sample, the FWHM shows no dependence with IP. In both objects, some 
CLs have widths comparable or even smaller than those displayed
by low-ionization lines, confirming previous results reported by \citet{b37}. 
All other remaining objects do not show a trend or it is difficult to 
evaluate it because all emission lines are unresolved. 

Table~\ref{tablex2} and Figures \ref{figpi} and \ref{figu} allow us
to state that overall, no particular trend between FWHM and IP is seen in regard
to the type of AGN. The subset of galaxies that we have labeled
``P'' contains similar proportion of both types of objects. Similarly, for
the objects where no trend in seen, no difference between Ty1 and Ty2 is
observed, reinforcing the results already discussed in the preceding
section. That is, a CLR located outside the torus region. Differences 
observed from object to object
is rather due to variations in the physical properties rather than
to a viewing angle dependence.

The increase of FWHM with IP up to a peak  energy of about 300 eV is  
interpreted  as due to a combined effect of increasing electron density  
gradient towards the centre and  spatial extension of the CL gas. The critical 
density of [S\,{\sc viii}] (IP=280.9~eV) and that of [S\,{\sc ix}] 
(IP=328.2~eV), two of the most common lines in the sample, are the highest 
among the CLs  studied: $n_{e}$ = 4$\times10^{10}$ and  2.5$\times10^{9}$ cm$^{-3}$, 
respectively. Thus, their critical 
density sets up an upper limit for the $n_e$ in the coronal line region 
to be $< 10^9$ cm$^{-3}$. However, the CLs  with higher IP (i.e.,
[Si\,{\sc  x}], [S\,{\sc  xi}] and [Fe\,{\sc  xiii}])  
have critical densities lower than the above values, by almost 
an order of magnitude (see Table~\ref{tab2}). Therefore, if the density 
increases steeply towards the center to values $\geq 10^8$ cm$^{-3}$, these  
later two high IP lines might get severely suppressed. Specifically, their  
high-velocity component would not be observed due to collisional de-excitation. 
Only their low velocity component - which should arise further out 
from the center -  would be seen. This would explain the observed 
decrease or constancy  of the line FWHM with IP.  

It is difficult to ascribe 
the absence of high velocity component in high IP lines to dust: absorption of 
the ionising photons will affect low and high velocity gas components alike.   
Depletion of the gas phase on dust is also not realistic as  that   
will tend, if any,  to reduce the low velocity component of the gas,  
which is formed further out from the dust sublimation radii, instead of the  
high velocity one (see Ferguson et al. 1997). In this respect, silicon  
lines, which should be largely depleted, show [Si\,{\sc x}]\,1.43\,$\mu$m 
(IP=351 eV) with FWHM comparable to that of [Si\,{\sc vi}]\,1.963\,$\mu$m 
(IP =167 eV) but still smaller than that measured for [S\,{\sc viii}] 
(IP=280) which is little affected by depletion. Clear examples are 
NGC\,262,  MCG\,05-13-17, ESO\,428-G014, Mrk\,1210, NGC\,4151, Mrk\,766, 
NGC\,5548 and NGC\,7469.

The results above allow us to conclude that {\rm (i)}
a strict limit for the density in the interface region between the  
coronal and the broad line region can be set in the range 10$^8 < n_e <  
10^9$ cm$^{-3}$; {\rm (ii)} the expected FWHM-IP correlation is modulated 
by the critical density of the corresponding ion. If a wide range of 
densities is present in the NLR - CLR region, we may indeed expect a strong  
correlation between the FWHM and the critical density, as already put  
forward by \citet{b16,b4,dro86} and \citet{fer97}.

\section{The relation between the coronal and  X-ray emissions}  \label{X-ray}

In order to produce the observed CLs, ionizing  energies in the range 127$-$450 eV 
are necessary. If photoionization by a central source is the dominant excitation 
mechanism, a positive relation between
X-ray strength  and the coronal line luminosity is expected.  Attempts have been 
made in order to look for connections  between X-rays  and  
CLs intensities but with conflicting results. \citet{b4} found no  correlation between 
[Fe\,{\sc  x}]~$\lambda 6374$ and the X-ray 2$-$10 keV luminosity, 
L$_{2-10 \rm{keV}}$. \citet{b60} 
found, based on numerous data extracted from literature, a clear correlation 
between the luminosity  of [Fe\,{\sc  x}]~$\lambda 6374$  and the ROSAT X-ray luminosity,
L$_{0.1-2.4 \rm{keV}}$. Similarly,
a positive trend, both in flux and luminosity, between the CLs   
[O\,{\sc  iv}]~25.9\,$\mu$m and [Ne\,{\sc  v}]~14.3\,$\mu$m  
and the ROSAT  L$_{0.1-2.4 \rm{keV}}$ emission is  found by \citet{b61}. 
However, these authors found no correlation when considering the hard X-ray 2$-$10 keV
luminosity. 

Considering the large dataset gathered in this work, the possible
relationship between the soft and/or the hard X-ray and the CL emission is
explored in more detail. We choose [Si\,{\sc  vi}]~1.963\,$\mu$m  
and  [S\,{\sc  viii}]~0.991\,$\mu$m 
as representative lines of the coronal region as they are the most frequent 
in the sample. Moreover, this comparison allows us to see if there are
differences between coronal lines emitted by refractory and non-refractory 
elements. 

Figures \ref{fig-xa} and~\ref{fig-xb}  display  
the soft X-ray luminosity, L$_{\rm x}$ [soft], integrated either over the 
0.1$-$2.4~keV or the 0.5$-$2.0~KeV interval, corrected by 
absorption, versus the luminosity of [Si\,{\sc  vi}]~1.963\,$\mu$m, 
and [S\,{\sc  viii}]~0.991\,$\mu$m, respectively. Only Ty1 sources
are included. The nature of the soft X-rays in Ty2 sources is questionable, in  
particular, in view of Chandra and XMM spectra, which show the X-ray emission  
resolved in emission lines \citep[e.g.][]{bgc06}. Under  
low level activity, a few Ty1 sources have also shown a soft X-ray  
emission line spectrum \citep[e.g.][and references therein]{nuc10}. 
Still, the bulk of their emission is expected to be  
dominated  by the AGN continuum light. 

It can be seen from Figs.~\ref{fig-xa} and~\ref{fig-xb} a trend of 
increasing CL $-$ with soft X-ray $-$ luminosity. The linear correlation
coefficient is 0.8 and 0.77, respectively, determined using
all points, with no distinction between normal and NLS1 galaxies.
The soft X-ray spans four orders of magnitude whereas the coronal
luminosity just two, so the  distance factor that stretch both axis  
should not be the major driver of the trend. These results
are consistent with the trend found by \citet{gel09} using
the luminosity of [Fe\,{\sc x}]~6374~\AA\ line and that of the 0.1-2.4 keV
band $L_x$.

Figures~\ref{fig-xe} and  
\ref{fig-xf} show the power law photon 
index, $\Gamma$, of the soft X-rays as defined by \citet{b59}
versus the coronal line luminosity for the same two lines. The
linear correlation indices in these cases is lower: 0.31 and 0.62,
respectively, very likely due to the scatter introduced by
NLS1 galaxies, which are usually characterized by a prominent soft excess.
Indeed, note that they have the largest values of $\Gamma$ and a higher
 CL luminosity.

To our knowledge, only \citet{b25} have explored  the incidence of 
$\Gamma$ on the CLs. They found that the values for the photon index 
are related to the  equivalent width of  [Fe\,{\sc  x}]~0.6375\,$\mu$m, 
concluding that strong CL emission is 
present predominantly in AGNs  with a soft excess ($\Gamma \geq$ 2.5).  
Figures~\ref{fig-xe} and  \ref{fig-xf} also suggest a photon index   
threshold: most objects  have $\Gamma$ above 2. NGC~3227 
has a harder spectrum (see both figures), meaning that it has a
deficiency in high energy EUV photons when compared to sources with
$\Gamma$ above 2. It is  
interesting to note that its CL spectrum has also a clear threshold in  
IPs, as CLs with IP $>$ 166\,eV are barely detected (see Fig.~\ref{fig-lay2}
and Sect 3).

Regarding the hard X-rays, Figure \ref{fig-y20} shows the absorption-corrected
luminosity, $L_{2-10\, {\rm keV}}$, versus the luminosity in [Si\,{\sc   vi}]~1.963\,$\mu$m.
For illustrative  purposes, Ty2 sources are now included in the figure. 
There is a large scatter in the plot but it may be noticed that this is  
introduced by the Ty2 sources mostly, normal Ty1s and NLS1s follow a  
narrower trend, with a linear correlation index of 0.97.
The number of objects with available hard X-emission  
is smaller, hence the reduced number of points in the figure as compared  
with those in Fig. \ref{fig-xa} or~\ref{fig-xb}. The same behavior is 
revealed when comparing with the [S\,{\sc viii}] luminosity - 
not shown.  The positive trends found reinforce the scenario of 
photoionization by the central source as the main mechanism for the 
formation of the coronal lines. The diagnosis fails however for Ty2 
sources because of their heavy absorption, even in the hard x-rays $-$some 
are Compton thick, e.g  NGC~1068, Circinus $-$ so that their measured 
X-ray luminosity is not accounting for the true intrinsic value.

The results above firmly support a relationship between soft and hard 
X-rays and NIR coronal lines,
explored for the first time in the literature here. They agree
with previous findings obtained in other wavelength intervals 
(optical and mid-IR) claiming a correlation between the luminosity of the
CLs and the X-ray  emissions (soft and hard) \citep{b60,b61,gel09}.
We should keep in mind that the X-ray and NIR observations 
are not contemporaneous but that should not affect the outcome. 
This is because, on average, the largest uncertainty comes from the 
model dependence inherent in converting X-ray count rates to fluxes 
rather than to variability itself \citet{gel09}. 

On the other hand, to be an X-ray AGN emitter is not condition to   
be also a CL emitter. Indeed, all the sources in the sample are X-ray  
emitters.  As discussed in Sect. 3, insufficient spatial resolution,  
specially for the more distant sources, hampers the  
detection of CLs.  However, that is not always the reason. A clear  
counter example is NGC~1097, one of the nearest AGN in the sample,   
with L(0.1- 2 keV) = 5.0~x~10$^{40}$  erg\,s$^{-1}$ \citep{ter02} and 
no CLs so far detected (Table 2), even  
at distances from the centre of 10~pc (Muller-Sanch\'ez et al. 2010, submitted). 
There are also AGNs with  a 
truncated CL spectrum at lines with IP above certain energy. Examples 
include NGC~3227, with barely detected CLs with IP $>$ 166 eV in the optical and 
IR (Rodriguez-Ardila et al. 2006, this work, see sect.3); Mrk~3, with 
no CLs with IP $>$ 100~eV within 10~pc from the centre \citep{m10}.
Some possibilities for the CL absence may include  
a particular type of ionizing continuum, i.e., one that is  
intrinsically different from that of an standard AGN, or is  
modified by intervening material located in between the central source  
and the CL region. Such material could be the  the warm absorber seen in  
some Seyfert galaxies \citep{np94,r97}, which could
significantly modify the form of the input ionizing continuum. \citet{b60} proposed that
this structure can also be a source of coronal lines. 
  
Another possibility  is that the
ionised gas in Ty2 galaxies have different distributions
in different sources \citep{s01a,s01b}. This hypothesis was raised in the 
light of {\it Chandra} observations
of Mrk~3 and Circinus. While in the former most of the line emission in X-rays is
spread over several hundred parsecs, this is not the case in the latter, where the
ionized gas is highly concentrated, with most of the flux originating within the
central 15~pc.

\section{Summary}  \label{con} 

The first study aimed at studying the physical properties of near-IR CLs 
spanning the energy range 130 $-$ 450 eV are derived for the first time 
on a sizable sample of AGNs (54 sources). The main results are as follows.

CLs in the 0.8$-$2.4~$\mu$m interval with IP $>$127~eV are detected in 2/3 (36 AGNs) 
of the galaxy sample. CLs are formed very close to the central source. 
Due to lost of line contrast in an increasing  continuum level, 
their detection depends largely on source distance. However, we confirm 
the existence of nearby objects, e.g. NGC~1097, NGC~5963, where the absence  
of CLs is genuine. These objects must have a suppressed  ionising continuum 
lacking photons with energies $\geq$ 100~eV.

Among the CLs and species studied in this work, from S, Si, Fe, Ca and Al, 
those coming from the the most abundant elements are the most frequent, 
namely Si and S. And among these, those from the lowest IP, 
i.e [Si\,{\sc vi}]~1.963$\mu$m (IP=166 eV) and [S\,{\sc viii}]~0.991 $\mu$mm (IP=280 eV)
are the most common ones. On the basis of these two lines, Ty2 sources 
with CLs outnumber Ty1 by $\sim$ 17\%. However, this discrepancy is 
attributed to differences in $z$: Ty1 sources have significantly 
larger $z$ than Ty2 and the two lines fall in regions of poor atmospheric 
transmission when $z~>$ 0.035. Focusing on the highest ionization lines, which are believed 
to be formed at the innermost regions of the AGN, their presence
in both types of AGNs in similar proportions (Fig. 8) implies that
their spatial location has to be outside the inner face of the torus. 
Thus, orientation effects do not play a role in the detection of coronal lines.

The luminosity of the coronal emission is distributed in a narrow interval, 
covering 3 dex at most. Moreover, the luminosity distribution of the four 
most frequent coronal lines studied ([Si\,{\sc vi}], [S\,{\sc viii}], 
[Si\,{\sc x}] and [S\,{\sc ix}]) is strongly peaked, with the largest number of
objects located between Log $L \sim$ 39 and $\sim$ 40 erg s$^{−1}$ cm$^{−2}$. 
[Si\,{\sc vi}]~1.963 $\mu$m is the most luminous of the coronal lines analyzed, 
confirming previous studies that use it as an indicator of nuclear activity.

A strong prediction of photoionization models, assuming a
simple relation between the cloud distance and its velocity, is a correlation  
between the line FWHM and the IP of the corresponding ion.  
We see a trend  of increasing FWHM with IP but up to  
energies of about 300 eV (Table 4). Above this peak energy, the FWHM
remains constant or decreases with IP. We ascribe this effect to a combination of
increasing electron density  towards the center  
and the spatial extension of the coronal gas. In a photoionized gas,  
the highest the IP the closer to the center a line is produced.
Coronal lines with the highest IP have critical densities
approaching 10$^8$ cm$^{−3}$. If the density in the innermost region of 
the NLR is significantly higher than that, those lines will be severely
affected by collisional de-excitation and their high-velocity component will 
get suppressed. Accordingly, only their lower velocity component, which 
might come from a further region from the center, would be seen.
This would explain why the FWHM of [S\,{\sc viii}]
is the highest for most sources: it is the coronal line with the highest 
critical density. Its intensity is not affected by collisional de-excitation 
as the lines of higher IP likely are. This scenario sets a strict limit  
for the density in the boundary region between the narrow and  the  
broad regions of 10$^8 - 10^9$ cm$^{-3}$.

Further support to photoionization as the main mechanism to excite
the CLs arises from the positive trends found between the X-ray luminosity 
$-$soft and hard, the soft power-law photon index and the coronal line luminosity. 
In particular, objects with the softest X-ray spectrum tend to have   
strong CLs, this being expected considering that the CLs surveyed in  
this work require energies in the 0.1 - 0.5 keV interval.
These trends hold  when considering Ty1 sources only;   
they get weaker or vanish  when including Ty2 sources. We believe this
is because the X-ray emission measured in these cases is not  
representative of the intrinsic AGN power. Some of the Ty2 sources 
are, indeed, Compton thick so even their 2-10 Kev spectrum may just be reprocessed 
emission. Other sources, like circumnuclear 
star formation, line emission and bremstrahlung are dominating the emission 
even in the hard X-rays.

The presence of truncated CL spectra in some objects, i.e., lacking CLs 
above a certain IP, could be ascribed to a specific geometrical
configuration, where the line of sight between the NLR gas and the soft X-ray continuum
is blocked by material that absorbs photons above a certain energy 
(a warm absorber, for example). It may also be possible that centrally illuminated 
ionized gas in Seyfert 
galaxies can have very different distributions in different sources. 
Alternatively, these  objects may have an special type of ionizing spectrum, 
lacking  ionizing photons above a certain energy threshold.
However, neither the corresponding  X-ray spectral index nor the X-ray luminosity
differentiate objects with and without CLs because they both share similar values 
in these two properties.

\acknowledgments

\noindent This research has been supported  by  the Instituto Colombiano para el 
Desarrollo de la Ciencia y la Tecnolog\'{\i}a, Francisco Jos\'e de Caldas,  (Colciencias),  
code 1101-05-17607, contract RC 206-2005 and by the Brazilian agency CNPq (308877/2009-8) 
to ARA. We are grateful to an annonymous Referee for a careful and critical reading of 
the manuscript.



\begin{figure}    
\plotone{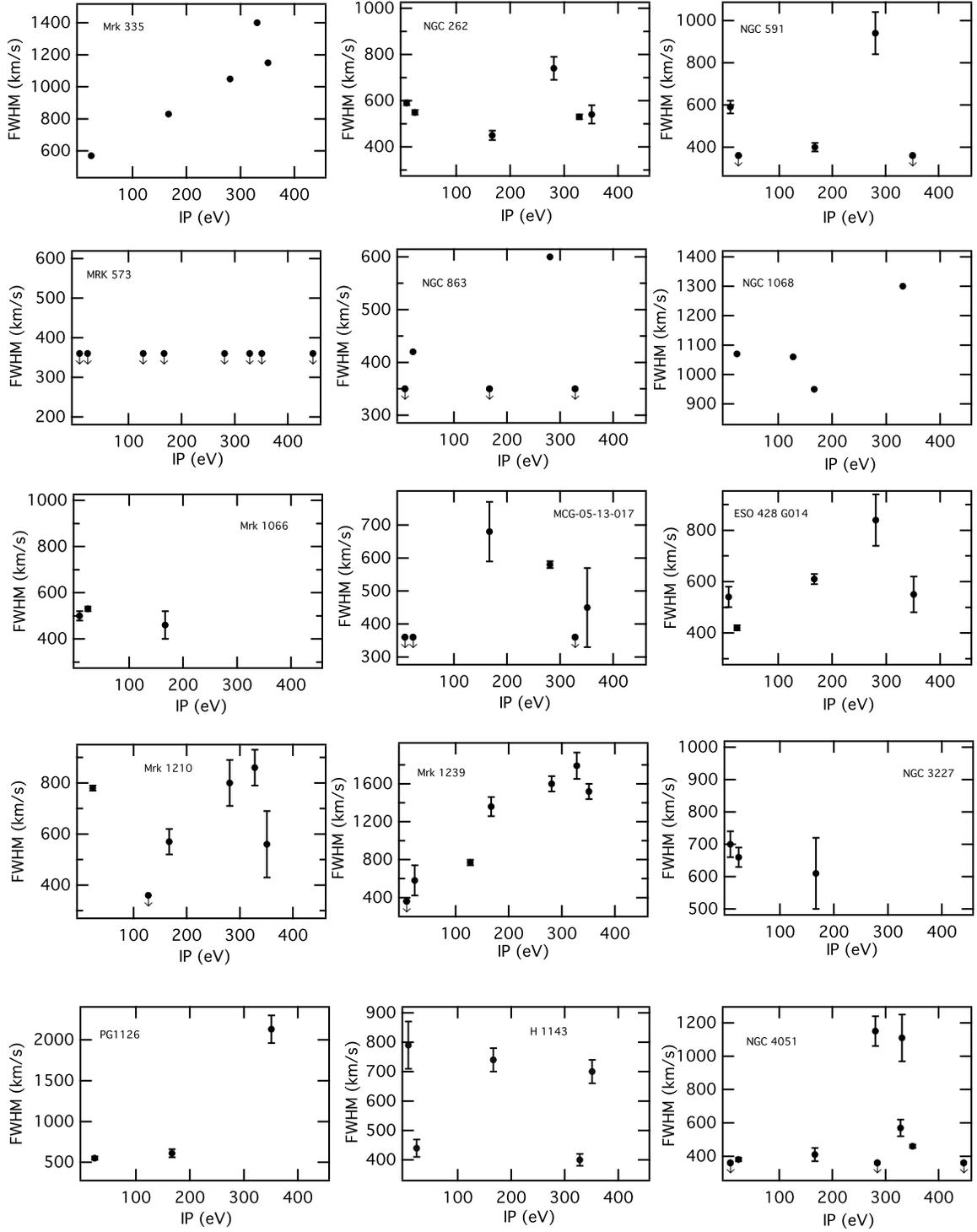}  
\caption{The FWHM vs IP for coronal lines detected in each galaxy. FWHM values are
corrected for instrumental resolution.  For purposes of 
comparison we include  the  low-ionization lines [Fe\,{\sc  ii}]\,1.2569~$\mu$m and  
[S\,{\sc  iii}]\,0.953 ~$\mu$m. Data points lacking error bars were taken from 
the literature, with no report of uncertainty.
 \label{figpi} }  
\end{figure}

\begin{figure}    
\plotone{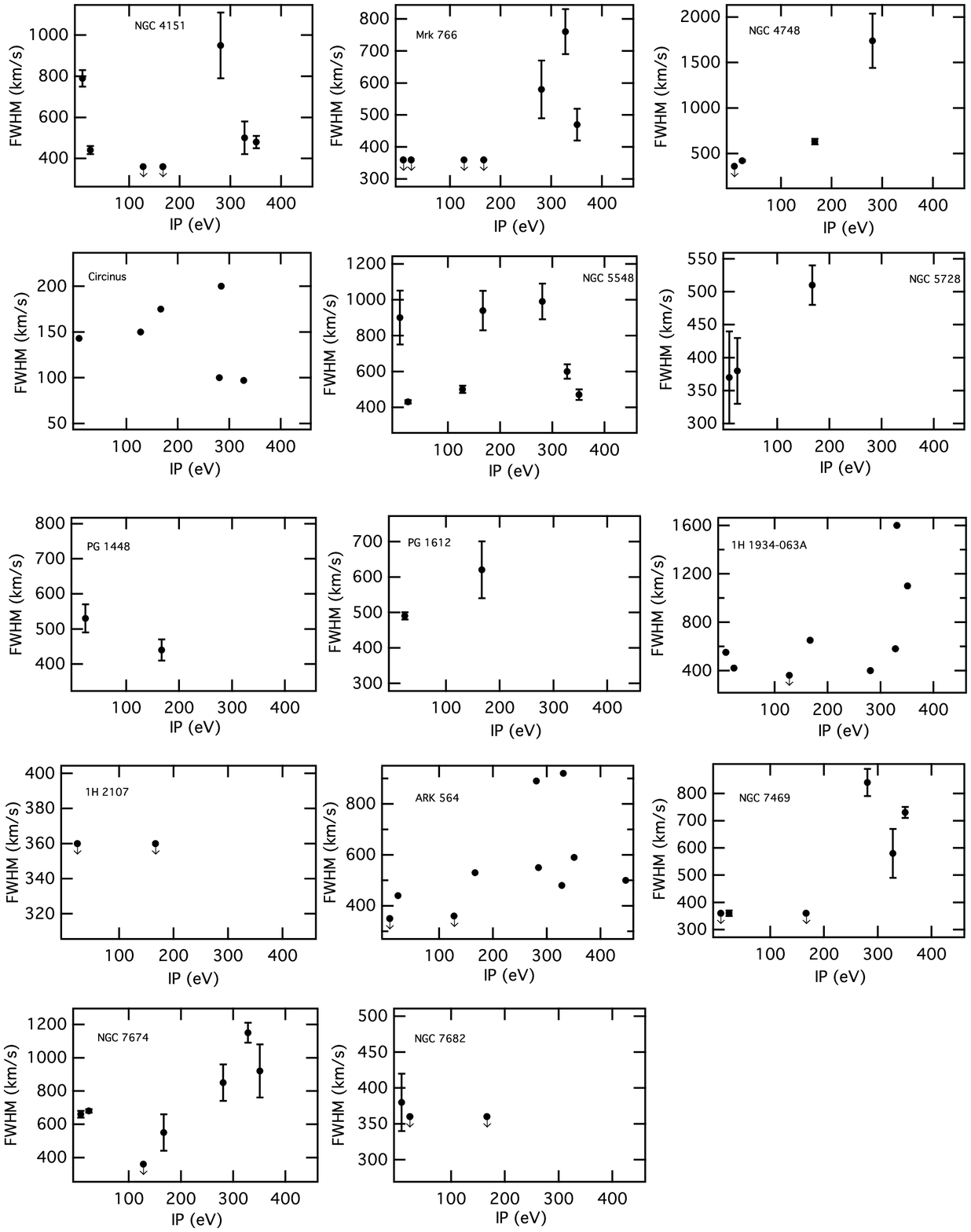}   
\caption{Cont. of Fig. ~\ref{figpi}.  \label{figu}  }   
\end{figure}

\begin{figure}    
\plotone{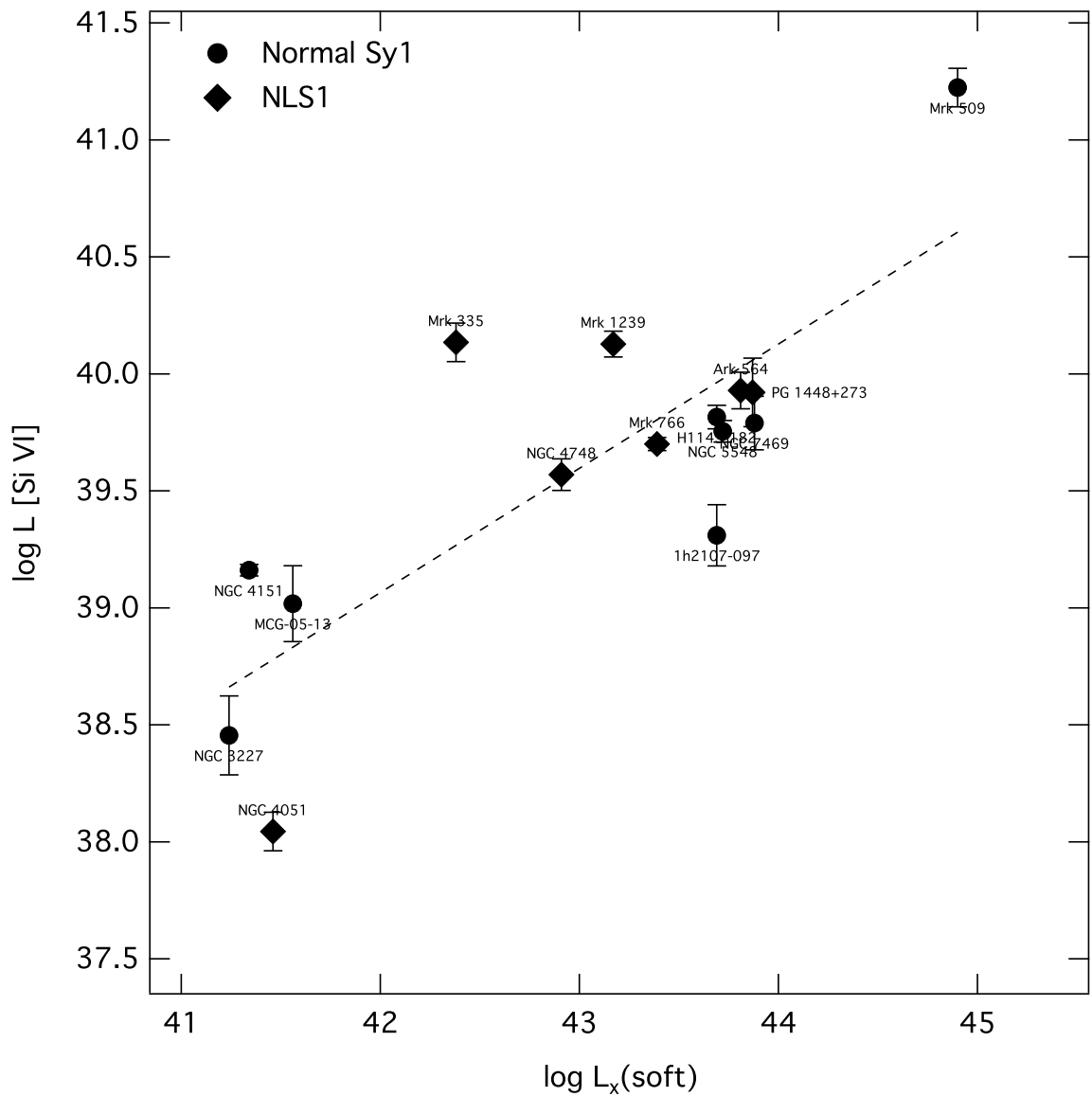}   
\caption{Luminosity of [Si\,{\sc vi}]\,1.963\,$\mu$m  (erg\,s$^{-1}$) versus the 
soft X-ray luminosity (erg\,s$^{-1}$). Soft X-ray data are from \citet{b57} (NGC\,3227, 
NGC\,4151, NGC\,5548, Mrk\,1239, NGC\,4051, Mrk\,766, NGC\,4748, NGC\,7469, Mrk\,493 and Mrk\,335), \citet{b58} 
(MCG-05-13-017, NGC\,7682 and Mrk\,509) and \citet{b59} (Ark\,564, PG1448+273
and H1143-182). The dashed line is a linear correlation fitted to the data. \label{fig-xa}  }   
\vspace{0.1cm}
\end{figure}

\begin{figure}     
\plotone{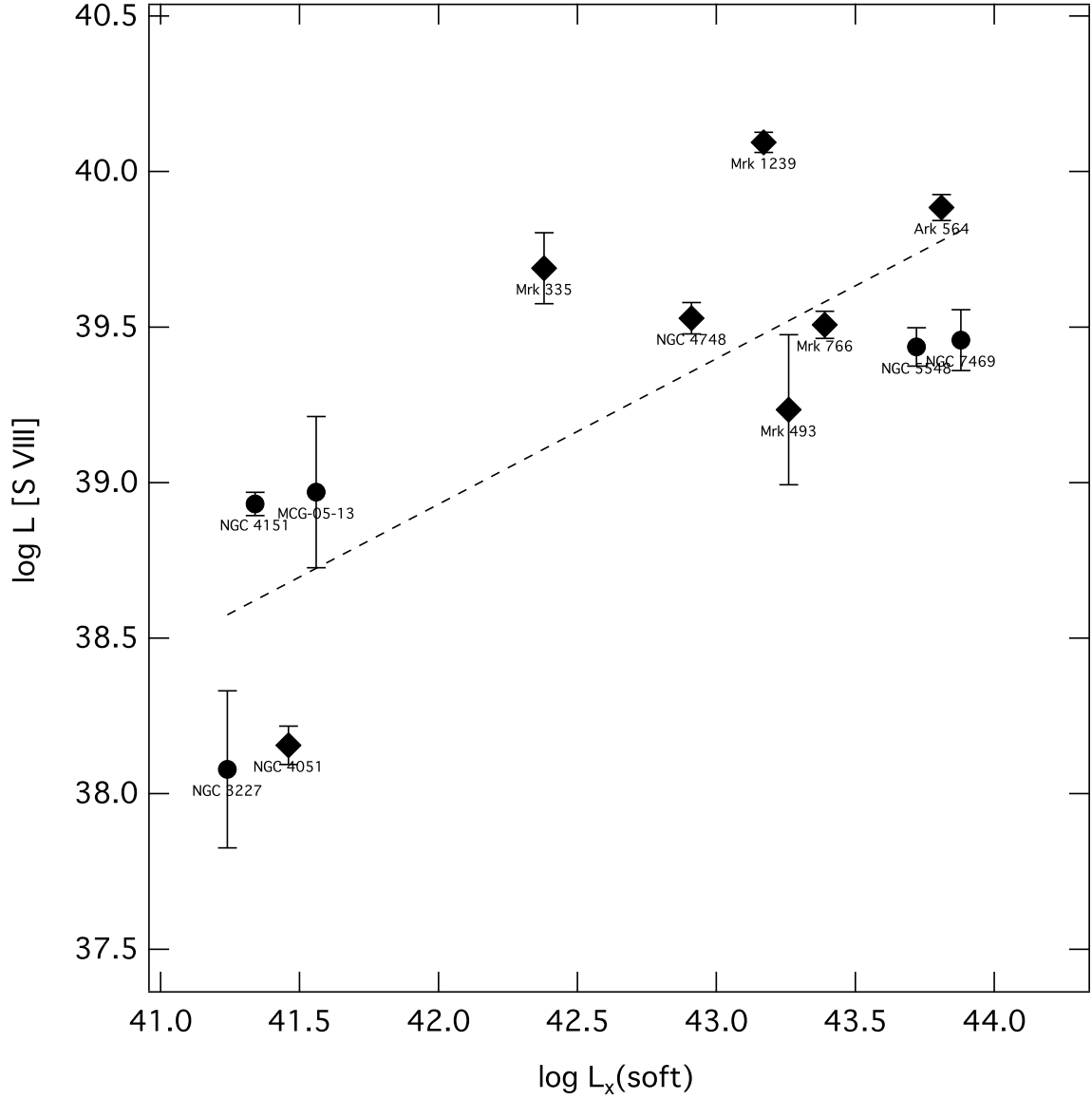}     
\caption{Luminosity of [S\,{\sc viii}]\,$0.991\,\mu$m  (erg.s$^{-1}$) vs the 
soft X-ray luminosity (erg.s$^{-1}$). Labels and  symbols are the same as in 
figure~\ref{fig-xa}. The dashed line is a linear correlation fitted to the data.
 \label{fig-xb} }    
\vspace{0.1cm}
\end{figure}

\begin{figure}    
\plotone{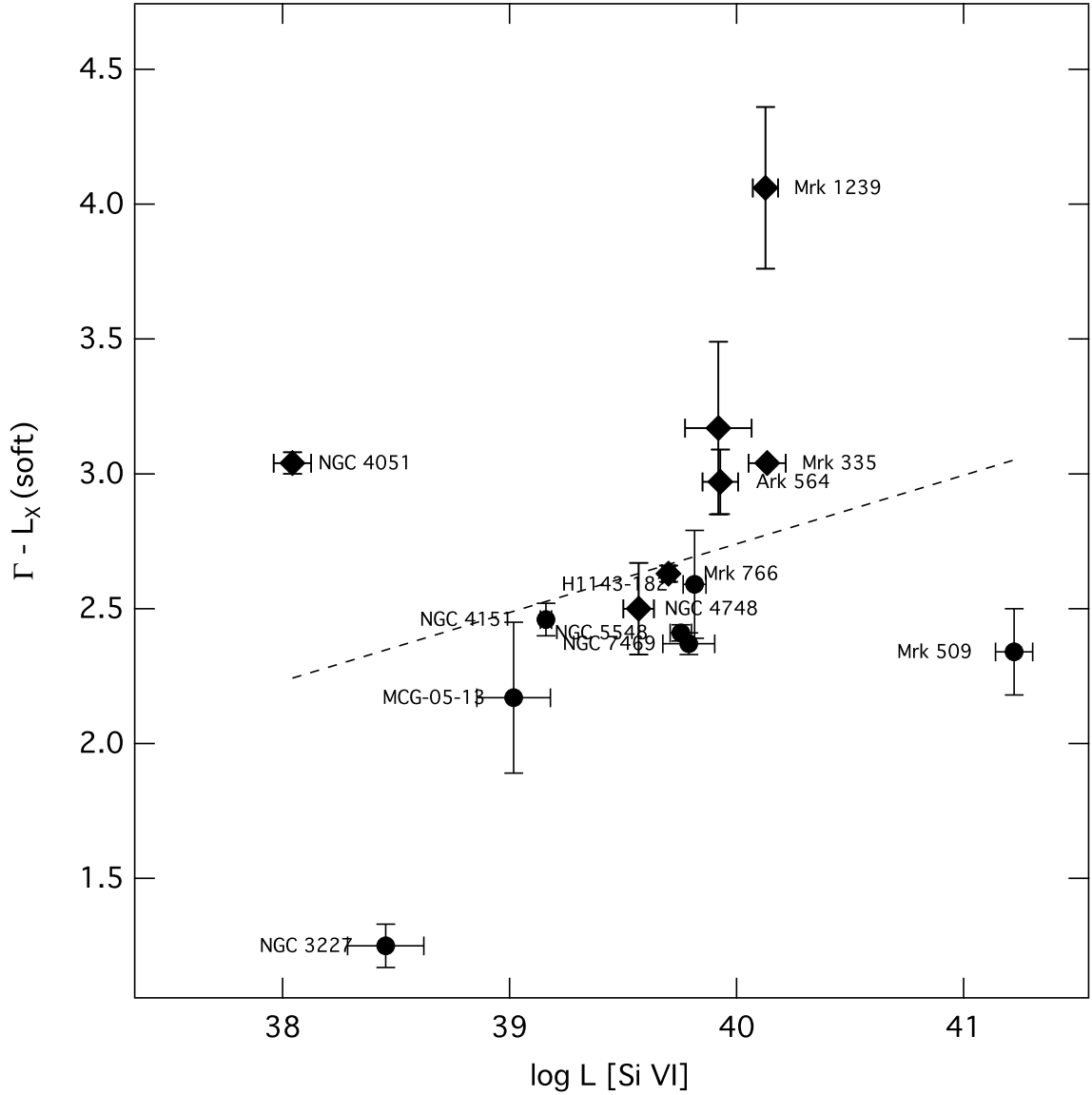}     
\caption{Photon index ($\Gamma$) of  soft  X-rays (0.1-2.4 keV) versus 
luminosity of [Si\,{\sc  vi}]\,$1.963\,\mu$m (erg.s$^{-1}$). All data of soft X-ray  
are from \citet{b57}. The dashed line is a linear correlation fitted to the data.\label{fig-xe}}    
\vspace{0.1cm}
\end{figure}

\begin{figure}   
\plotone{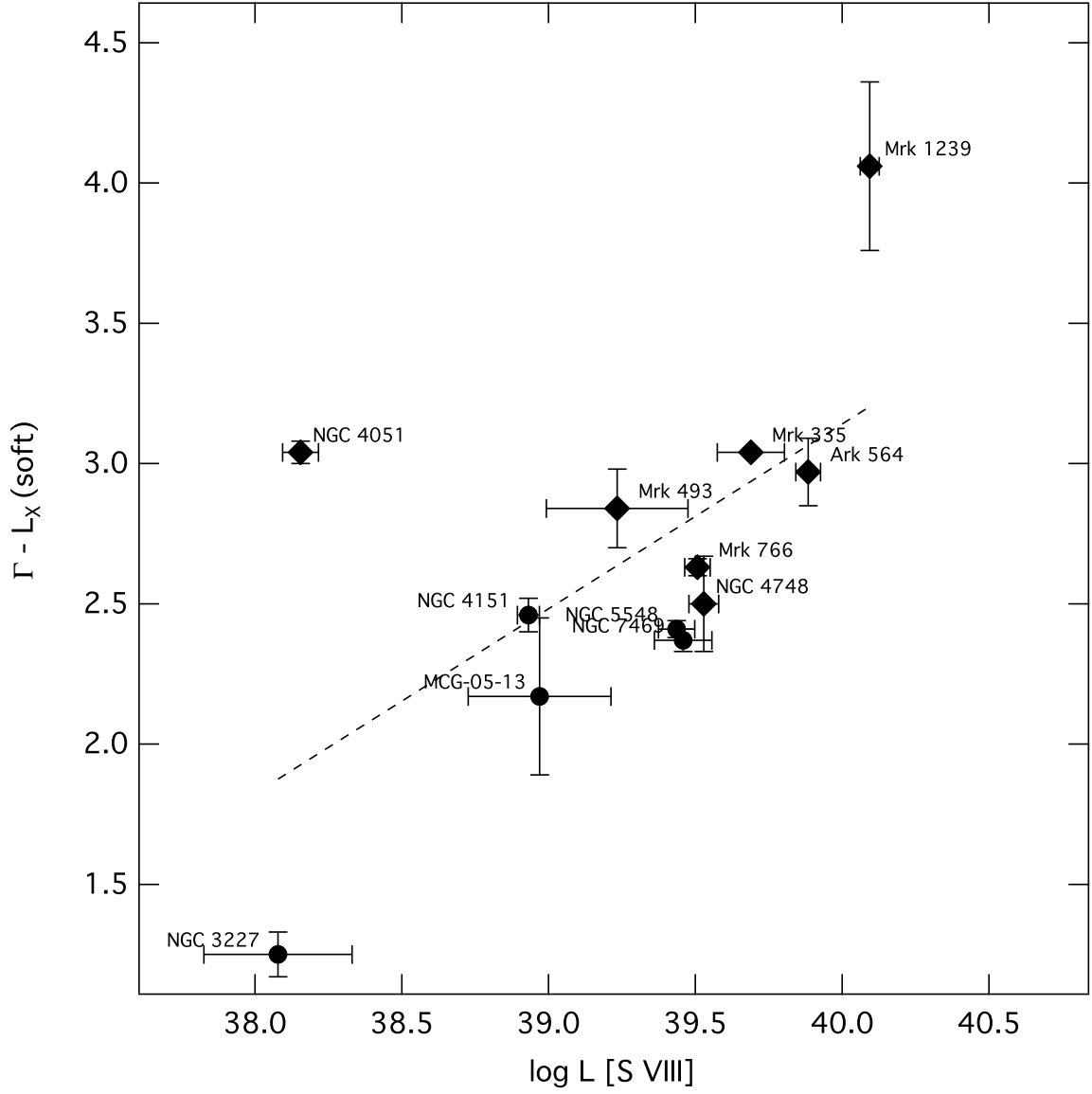}    
\caption{Same as in Fig. \ref{fig-xe}  but now in the abscissa is the luminosity of 
[S\,{\sc  viii}] $0.991\,\mu$m  (erg.s$^{-1}$).   \label{fig-xf}  }  
\vspace{0.1cm}
\end{figure}

\begin{figure}    
\plotone{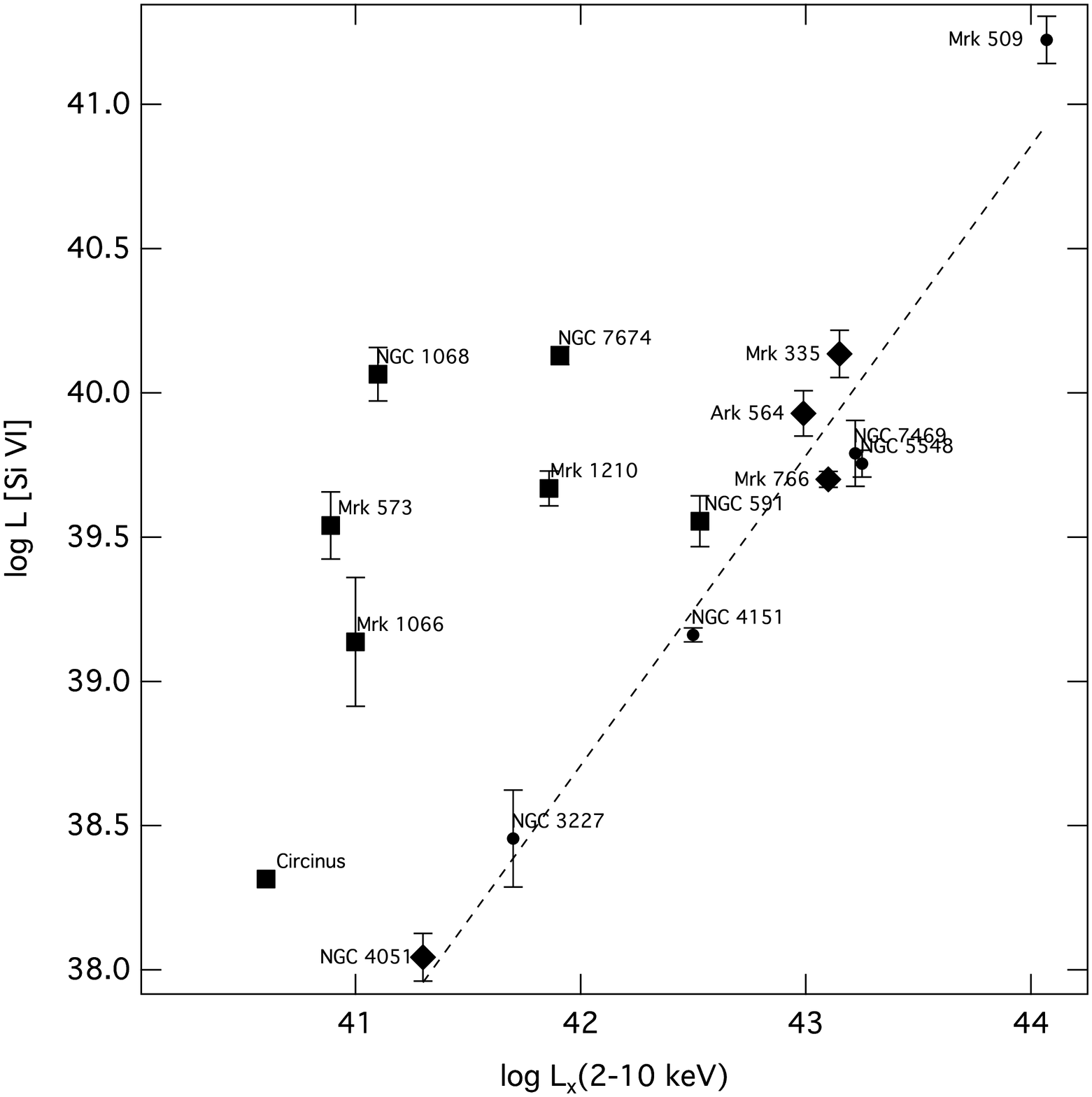}    
\caption{Luminosity of [Si\,{\sc  vi}] $1.963\,\mu$m  (erg.s$^{-1}$) versus 
the hard  X-ray luminosity (erg.s$^{-1}$).  Labels have the same meaning as in 
Fig. \ref{fig-xa}.  Data of hard X-ray are from
\citet{b77} (NGC\,3227, NGC\,4151, NGC\,4051, NGC\,1068); \citet{b84} (NGC\,7674, 
Mrk\,1210,  Circinus); \citet{b80} (NGC\,7469,  Mrk\,509);  \citet{b83} (Mrk\,573, 
NGC\,591); \citet{b86} (Mrk 335, Mrk 766); \citet{b81} (Ark 564); 
\citet{b82} (Mrk 1066)  and \citet{b79} (NGC 5548).
The dashed line is a linear correlation fitted to the data.   \label{fig-y20}  }   
\vspace{0.1cm}
\end{figure}

\begin{deluxetable}{llrrr}   
\tablewidth{0pt}
\tablecaption{Galaxy sample. Redshifts were taken from NASA/IPAC Extragalactic 
Database (NED). The last column indicates if the correspondent galaxy emits at 
least one coronal line (Y) or none (N) with IP $\gtrsim$100~eV in its NIR spectrum .  
\label{tab1}}.     
\tablehead{
\colhead{Galaxy}  &   \colhead{Type} &   \colhead{$z$}  &  \colhead {$\arcsec$/pc} & \colhead{CL}}  
\startdata
Mrk 334                             &    Sy1   &  0.021940  & 460     &   N$^{\textrm {a}}$   \\ 
Mrk 335*                           &   NLS1  &  0.025785  & 500         &   Y      \\
NGC 262                          &    Sy2   &  0.015034 & 290           &   Y    \\  
Ton S180*                        &   NLS1  &  0.061980  & 1200         &   N  \\
Mrk 993                            &    Sy2 &   0.015537  & 300        &   Y            \\  
NGC 591                        &     Sy2   &  0.015167 & 290      &   Y       \\  
Mrk 573                           &     Sy2 &  0.017179 & 330      &   Y    \\  
NGC 863*                       &     Sy1   &  0.026385  & 550        &   Y      \\ 
Mrk 1044*                       &    NLS1       &  0.016451 & 320     &   N      \\    
NGC 1068*                     &    Sy2   & 0.003793  & 75  &   Y        \\
NGC 1097                      &    Sy1 &  0.004240   & 90      &   N           \\ 
NGC 1144                      &    Sy2   &  0.028847   & 560       &   N$^{\textrm {a}}$   \\   
Mrk 1066                        &    Sy2  &  0.012025 & 230        &   Y          \\ 
NGC 1275                      &    Sy2  &   0.017559  & 340     &   N          \\  
MCG-05-13-017            &     Sy1  &  0.012445 & 260       &   Y        \\ 
NGC 2110                      &     Sy2 &  0.007789   & 150      &   N$^{\textrm {a}}$  \\  
ESO 428-G014              &    Sy2   &  0.005664 & 110     &   Y    \\   
Mrk 78*                           &    Sy2  &   0.037150   & 720    &   Y  \\ 
Mrk 1210                        &    Sy2    &   0.013496   & 260    &   Y   \\
Mrk 124                          &   NLS1 &   0.056300  & 1090           &   Y   \\ 
Mrk 1239                        &    NLS1 &   0.019927  & 380           &   Y      \\ 
NGC 3227                      &    Sy1  &   0.003859   & 80    &   Y    \\  
PG 1126-041                 &    NLS1 &   0.060000  & 1160    &   Y  \\
H1143-182                    &   Sy1  &   0.032949  & 680     &   Y   \\
NGC 4051                     &   NLS1 &   0.002336  & 45       &   Y      \\ 
NGC 4151                     &   Sy1  &  0.003319  & 70       &   Y  \\ 
Mrk 766                          &   NLS1 &   0.012929 & 250      &   Y    \\  
NGC 4748                     &   NLS1  &   0.014630  & 280         &   Y   \\  
Ton 0156                       &   QSO   &   0.549000  & 11400       &   N  \\
Mrk  279                         &   NLS1   &   0.030451  & 590       &   Y   \\ 
Circinus*                        &   Sy2 &   0.001448   & 30        &   Y      \\
NGC 5548                     &   Sy1 &   0.017175   & 360      &   Y    \\ 
PG 1415+451               &   Sy1   &    0.113587   & 2350       &   N  \\
Mrk 684                         &   Sy1  &   0.046079 & 950          &   N   \\
Mrk 478                         &   NLS1   &  0.079055 & 1530           &   N  \\
NGC 5728                    &   Sy2   & 0.009353   & 180       &   Y  \\  
PG 1448+273              &    NLS1 &  0.065000  & 1250        &   Y \\
PG 1519+226              &    NLS1 &   0.013700 & 2650         &   N        \\
NGC 5929                    &    Sy2  &  0.008312  & 160        &   N$^{\textrm {a}}$  \\   
NGC 5953                    &   Sy2  &  0.006555  & 130      &   N    \\   
Mrk  291                       &    NLS1 &   0.035198  & 680    &   N    \\
Mrk 493                        &    NLS1 &   0.031328  & 610    &   N$^{\textrm {a}}$  \\
PG 1612+261              &    Sy1 &  0.130916  & 2700      &   Y      \\
3C 351                          &   QSO &  0.371940  & 7700    &   N  \\
Mrk 504                        &    NLS1  &   0.035888 & 700           &   Y   \\
Arp 102B                      &   Sy1  &    0.024167  & 500    &   N   \\ 
1H 1934-063A            &  NLS1    &  0.010254   & 200 &  Y  \\ 
Mrk 509                       &   Sy1  &   0.034397  & 710      &   Y   \\ 
Mrk 896                       &   NLS1       &   0.026424  & 510    &    Y  \\ 
1H 2107-097              &   Sy1   &  0.026525  & 550      &   Y  \\
Ark 564                        &   NLS1       &    0.024684 & 480     &   Y   \\ 
NGC 7469                   &   Sy1   &   0.016317 & 340     &    Y     \\ 
NGC 7674                   &   Sy2  &   0.028924   & 560    &   Y         \\  
NGC 7682                   &   Sy2  &  0.017140   & 330    &   Y     \\  
\enddata
\tablenotetext{*}{Data taken from the literature.} \\
\tablenotetext{a}{No coronal lines were detected in the 0.8$-$2.4\,$\mu$m 
region but other authors have reported coronal lines in other
spectral regions. See Sect~\ref{obs} for further details.}
\end{deluxetable}

\clearpage

\begin{deluxetable}{lcccccccccc}    
\tabletypesize {\scriptsize}
\rotate 
\tablecaption{Fluxes of the CLs for the galaxy sample in units of 
$10^{-15}$ erg cm$^{-2}$\,s$^{-1}$. For the recombination lines the 
fluxes correspond to the narrow component, 
except when specified. \label{tab70}}
\tablewidth{0pt}
\tablehead{
      &    \colhead{Pa$\alpha$} &  \colhead{Br$\gamma$} &  \colhead{[Ca 
VIII]} &   \colhead{[Si VI]} &   \colhead{[S VIII]}  &   \colhead{[Al 
IX]}  &  \colhead{[S IX]} & \colhead{[Fe XIII]} &  \colhead{[Si X]} & 
\colhead{[S XI]} \\ 
     Galaxy  &    \colhead{1.8750 $\mu$m} &  \colhead{2.1654 
$\mu$m}        &    \colhead{2.3213 $\mu$m}  &  \colhead{1.9630 $\mu$m}  
&  \colhead{0.9913  $\mu$m}   &       \colhead{2.0450 $\mu$m}   &  
\colhead{1.2520 $\mu$m}  &        \colhead{1.0747 $\mu$m}  &      
\colhead{1.4301 $\mu$m}  &     \colhead{1.9196 $\mu$m}}  
\startdata
Mrk 334     &  $12.6\pm0.4$  &    $4.7\pm0.2$     &     ***             
&            $<2.0$              &      $<0.7$           
&         $<0.8$                 &     $<1.4$           
&           $<1.8$                         & $<1.8$    &  
$<1.3$    \\
Mrk 335&  ...  &   $26.7\pm3.1\S^{\textrm{a}}$   &    $\cdots$ 
&$10.6\pm 2.0^{\textrm{a}}$ &  $3.8\pm1.0^{\textrm{a}}$  &  $\cdots$ & 
$12.6\pm 2.5^{\textrm{a}}$ & $18.3\pm 3.8^{\textrm{a}}$ & $7.5\pm 
3.0^{\textrm{a}}$ & $\cdots$ \\
NGC 262    &  $35.3\pm1.1$  &   $1.1\pm0.1$       &    $<0.8$    
&    $4.4\pm 0.7$    &    $2.7\pm 1.1$    &    $<0.3$     
&    $2.3\pm 1.1$    &    $<1.7$     &    $2.0\pm 1.1$    
&    $<1.8$     \\
Ton S180   &  ...  &   $5.3\pm1.0\S^{\textrm{a}}$       &       
*       &         $\cdots$            &         $\cdots$               
&      $\cdots$       &                $\cdots$        &      
$\cdots$          &   $\cdots$    & $\cdots$   \\
Mrk 993   &  **  &   $\cdots$         &      ***             
&            $<2.0$                                                  
&        $<0.9$                                             
&            $<0.5$        &         $<1.1$             &        
$<1.8$    &   $<1.7$ & $<1.8$    \\
NGC 591    &  $60.1\pm0.8$  &  $3.6\pm0.2$  &    ***    &    $8.1\pm 
1.6$    &    $3.3\pm 1.4$    &    $<0.7$     &    $<1.7$ 
    & $<2.1$     &    $3.2\pm 1.6$    &    $<1.6$    \\
Mrk 573    &  $45.5\pm0.3 $  &   $2.8\pm0.1 $        &    $3.2\pm 
0.5$    &    $6.1\pm 1.6$    &    $6.1\pm 0.9$    &    
$<0.6$  &     $6.2\pm 0.9$    &    $3.3\pm1.6$     &    
$7.5\pm 1.5$    &    $2.2\pm1.7$    \\
NGC 863    &  ...  &    $9.3\pm2.6\S^{\textrm{a}}$       &    
$\cdots$     &    $<2.1^{\textrm{a}}$    &    $<1.0^{\textrm{a}}$    
&    $\cdots$     &    $1.1\pm 0.4^{\textrm{a}}$    &$\cdots$ &    
$3.2\pm1.0^{\textrm{a}}$        &    $\cdots$     \\
Mrk 1044  &  ...  &   $7.8\pm1.5\S^{\textrm{a}}$       &     
$\cdots$      &     $\cdots$      &           
$\cdots$                                &    $\cdots$     &            
$\cdots$                                        &     $\cdots$        &  
**      &   $\cdots$    \\
NGC 1068  & $\cdots$ &  $34.6\pm4.1^{\textrm{k}}$     &    
450$^{\textrm{b}}$    &    $421\pm90^{\textrm{c}}$    &    
$87^{\textrm{b}}$    &    $\cdots$     &    $50^{\textrm{b}}$    &    
...   &    $270\pm50^{\textrm{e}}$    &    $\cdots$     \\
NGC 1097   & **   &   $\cdots$      &     ***     &   $<3.8$     &   
$<2.8$    &    $<0.9$     &    $<3.0$     &       
$<3.0$         &    $<2.5$                  &  **     \\
NGC 1144  & $\cdots$&$\cdots$& *** &  $<0.9$     &       
$<0.6$     &    $<0.2$     &  $<0.4$       &  
$<1.2$      &     $<0.5$          & $<0.8$      \\
Mrk 1066   &  $145.7\pm1.6$   &    $14.2\pm0.2$    &     ***    &    
$4.9\pm 2.5$    &    $<1.9$     &    $<1.1$     &    
$<2.2$    &    $<2.1$     &    $<2.9$     & $<3.4$     \\
NGC 1275  & $145.1\pm2.5$   &   $9.8\pm0.4$       &     
$<1.2$        &      $<1.5$       &    $<1.4$      &    
$<0.9$      &     $<2.5$        &   $<1.9$    &   
$<1.2$        &   $<0.7$    \\
MCG-05-13-017 &  **  & $0.8\pm0.1\S$  &    ***    &    $3.5\pm 
1.3$    &    $3.1\pm 1.7 $    &    $<0.9$     &    $2.0\pm 
1.6$ & $<1.5$ &$3.7\pm 1.6$  &  $<1.7$     \\
NGC 2110  &  $33.0\pm4.5$  &    $2.5\pm0.2$     &    
$<0.6$               &    $<0.6$    &    $<1.1$     &    
$<0.9$     &       $<1.0$    &    $<4.0$     &  
$<5.0$        &   $<1.6$       \\
ESO 428-G014     &  $102.1\pm0.8$  & $9.0\pm0.1$   &    ***     &    
$21.1\pm 3.9$    &    $4.0\pm 2.8$    &     $<1.8$       
&     $<3.8$       &     $<3.8$       & $4.2\pm 2.2$       
&     $<3.7$   \\
Mrk 78 &  $17.5\pm2.5^{\textrm{f}}$  &   $2.5\pm0.2^{\textrm{f}}$  &    
 $\cdots$    &    $19.7 \pm 1.7^{\textrm{f}}$    &    $1.7 \pm 
0.3^{\textrm{f}}$    &     $\cdots$       &     $\cdots$       &    
 $\cdots$       &     $\cdots$       &     $\cdots$       \\
Mrk 1210& $57.2\pm1.4$   &   $4.6\pm0.2^{\textrm{g}}$ &    
$1.6\pm 0.1^{\textrm{g}}$    & $13.3\pm 0.9^{\textrm{g}}$ &    $2.9\pm 
0.2^{\textrm{g}}$    &     $<0.5$   &$4.2\pm 
0.5^{\textrm{g}}$    & $<1.8$  &$2.8\pm 0.4^{\textrm{g}}$    & 
$<2.7$       \\
Mrk 124 &  $15.2\pm0.5$  &      $2.5\pm0.6\S$     &      *       
&              $<0.4$                          &                 
$<0.5$         &        $<0.4$         &       $<0.8$      
&     $<2.1$         &    $<0.7$         &   $<0.3$        \\
Mrk 1239     &  $50.5\pm2.7$  &   $14.0\pm2.1$       &    $5.7\pm 
1.3$    &    $17.5\pm 2.2$    &    $16.2\pm 1.2$    &    
 $<1.1$       &    $15.5\pm 1.8$    & $<3.5$   &    $26.2\pm 
1.9$    &     $<2.8$       \\
NGC 3227  &  $71.7\pm2.8$      & $6.5\pm1.3$ &    $<0.9$    &    
$10.0\pm 3.9$    &    $4.2\pm2.4$    & $<1.1$   &    
 $<2.2$       &     $<2.6$       &    $<3.9$    &    
 $<3.4$       \\
PG 1126-041 & $32.8\pm1.1$   & $13.3\pm0.7\S$ &     *       &    
 $4.1\pm0.4$       &    $<0.8$    &     $<0.5$       &    
$<1.3$    & $<2.8$       & $5.1\pm0.7$  & $<0.8$    \\
H1143-182     &    $106.6\pm1.4$  & $10.96\pm0.7$  &    
$<0.9$         &       $3.1\pm3.6$    &    $<1.4$  &      
$<0.2$        &            $1.0\pm0.5$    &     
$<1.1$        &    $1.6\pm0.5$         &      
$<0.4$             \\
NGC 4051  & $111.5\pm4.3\S$  &  $13.1\pm0.9\S$ &    $2.4\pm 0.9$ & 
$10.6\pm 2.0$ & $13.7\pm 2.0$    &$1.8\pm1.1$&    $11.1\pm 
2.3$    &    $8.0\pm2.2$&$ 22.2\pm 2.5$    & 
$5.9\pm1.7$        \\
NGC 4151  &  $581.5\pm12.0\S\star$  &   $12.4\pm0.6$    &    $15.1\pm 
2.3$    &    $68.7\pm 3.9$    &    $40.5\pm 3.5$    &    
$2.3\pm1.6$&    $39.7\pm 2.3$& $<1.6$       &    $37.7\pm 
3.0$    &    $<10.6$         \\
Mrk 766    & $27.2\pm0.5^{\textrm{h}}$   &   $7.4\pm0.4^{\textrm{h}}$ & 
$2.5\pm 0.6^{\textrm{h}}$ &    $15.6\pm 1.0^{\textrm{h}}$ & 
$10.0\pm1.0^{\textrm{h}}$    &  $\cdots$    &    $9.4\pm 
1.0^{\textrm{h}}$    &  $\cdots$    &$6.5\pm0.6^{\textrm{h}}$  &    
 $\cdots$    \\
NGC 4748  &  $43.2\pm0.6$  & $1.4\pm0.1$      &    $<1.4$    &    
$9.0\pm 1.4$    &    $8.2\pm1.0$    &     $<0.8$       &    
 $1.9\pm0.7$       &     $<1.6$       &     $3.3\pm1.3$   
    & $<1.4$           \\
Ton 0156     &  *  &   *      &        *       &                     
*                                    &              $<0.2$          
&             $\cdots$                    &       $<0.4$        
&     $<0.4$       &   $<0.1$        &     *    \\
Mrk  279    &  $17.5\pm0.5\S\star$  &   $0.6\pm0.4$        &    
 $<0.8$                & $<0.8$      &     $<0.5$       &    
 $<0.4$       &     $<0.7$       &     $2.9\pm1.1$       
&     **       &     $<1.1$       \\
Circinus     &  $\cdots$  &  $14.1^{\textrm{i}}$  &    
48$\S^{\textrm{i}}$    &    51.5$\S^{\textrm{i}}$    &    
11$^{\textrm{j}}$    &     10.8$\S^{\textrm{i}}$       &    
20$^{\textrm{j}}$    &     $\cdots$       &     $\cdots$       & 
$\cdots$           \\
NGC 5548  &  **  &  $16.3\pm2.0\S$      &    $1.7\pm 0.8$    &    
$10.0\pm 1.1$    &    $4.8\pm 0.7$    &     $<0.3$       &    
$3.7 \pm 0.7$    &     $<0.6$       &    $5.6\pm 1.0$    
&    $1.5\pm1.2$     \\
PG 1415+561   &  $21.4\pm0.7\S$  & $\cdots$   &        *       
&           $<0.2$          &     $<0.4$         &       
$<0.1$       &                         **                            
&           $<0.4$          &   $<0.5$   &  $<0.2$         \\
Mrk 684     &  $24.3\pm0.5$  &    $4.5\pm0.5\S$      &         
*       &            $<0.6$     &            $<0.9$         
&      $<0.4$          &     $<0.5$       &     
$<3.6$          &    $<0.5$          &   $<0.7$       \\
Mrk 478    &  $31.2\pm0.6$  &   $7.0\pm0.6\S$ &        *         
&     $<0.3$        &    $<0.8$        &     $<0.4$       
&     $<1.4$         &        $<1.4$          &     
$<0.7$         &   $<0.3$      \\
NGC 5728  &  $20.6\pm1.9$   &  $2.1\pm0.2$     &     $<0.9$    &    
$7.1\pm 1.8$    &     $<1.7$       &     $<0.5$       &    
 $<2.0$        &     $<2.9$       &     $<1.2$       &    
 $<0.6$       \\
PG 1448+273  &  $27.0\pm1.2$  &  $1.5\pm0.2$    &    *    &    $1.0 \pm 
0.3$    &    $<0.8$    &     $<0.2$       &     $<1.1$   
    &     $<0.8$       &     $<0.3$       &     $<0.3$       \\
PG 1519+226  & $9.8\pm0.4$   &  *  &      *          &       
$<0.2$         &       $<0.6$       &      $<0.1$       
&     $<0.6$       &      $<0.5$       &     $<0.4$          
&   $<0.2$     \\
NGC 5929  &  **   &   $1.4\pm0.3$ &        $<0.5$              
&       $<0.9$         &    $<1.0$        &       
$<0.4$         &          $<0.9$       &      $<2.1$       
&      $<3.0$         &   $<5.6$      \\
NGC 5953   &  $19.3\pm0.8$  &    $\cdots$    &    ***    &       
$<3.8$        &      $<1.3$         &      $<1.1$       &    
$<3.4$        &    $<3.4$        &     $<3.8$         &  
$<2.6$      \\
Mrk 291   &  $9.4\pm0.2$  &  $0.4\pm0.1$    &     ***         &      
$<1.7$       &          $<0.3$        &     $<0.2$        
&        $<0.5$       &      $<1.0$           &    
$<0.5$          &  $<0.1$       \\
Mrk 493   & $23.8\pm0.5$   &  $1.8\pm0.3$   &      
$<0.5$                &     $<0.4$         &      
$0.9\pm0.5$       &      $<0.5$          &       
$1.4\pm0.8$        &    $<0.8$       &    $<0.9$         
&    $<0.4$      \\
PG 1612+261 &  $13.2\pm0.2$  &  * &    *    &    $2.4\pm 0.4$    
&    $<0.8$    &     $<0.3$       &    $<0.7$    &    
 $<0.4$       &    $<0.6$    &     $<0.3$  \\
3C 351   &  *  &  *    &        *         &                     
*                                       &                 
**                            &                  
*                                        &           $<1.0$          
&      $<2.0$        &     $<1.1$         &   *  \\
Mrk 504    &  $9.6\pm0.4$  &  $3.4\pm 0.5\S$  & $<0.2$  &    
$<0.2$    &    $0.5\pm0.3$     &     $<0.1$       &    
 $<0.6$       &    $2.2\pm0.8$    &    $<0.6$    &    
 $<0.1$       \\
Arp 102B   &  $7.1\S\star$  &   $\cdots$      &    $<0.5$       
&      $<0.9$       &      $<0.5$            &      
$<0.3$       &      $<1.2$                    &   $<1.6$     
&    $<1.1$      &   $<0.8$     \\
1H 1934-063A& $55.0\pm2.5$  & $22.4\pm0.1\S^{\textrm{a}}$ & 
$3.9\pm1.1^{\textrm{a}}$& $24.9\pm2.5^{\textrm{a}}$ 
&$12.3\pm2.8^{\textrm{a}}$& $\cdots$ &$9.0 \pm 
2.0^{\textrm{a}}$&$29.2\pm4.0^{\textrm{a}}$&$20.0\pm5.4^{\textrm{a}}$&$\cdots$ 
\\
Mrk 509   &  $1422.9\pm52.1$  &   $382.0\pm33.3\S$         &      
$<7.8$                &                 $72.7 \pm13.8$ 
&            $<11.9$           &     $<6.8$     &           
$<8.2$          &      $<12.9$         &    **      &  
$<16.3$     \\
Mrk 896     &  $2.7\pm0.3$  &   $0.4\pm0.1$  &  $<0.6$  &   
$<0.7$            &              $<0.8$        &             
$<0.2$        &         $<0.6$       &              
$<1.0$      &     $<0.7$        &  $<0.7$  \\
1H 2107-097  &  $71.7\pm2.0\star$  & $20.9\pm1.9\S$ &    
$1.4\pm0.4$&    $1.5\pm 0.4$    &    $<1.8$    &$<0.3$   
&    $<1.2$     &     $29.2\pm2.3$       &    $2.3\pm0.9 $ & 
$<0.6$       \\
Ark 564& $62.2\pm0.8$ & $10.5\pm0.6\S^{\textrm{a}}$ & $3.1\pm 
1.0^{\textrm{a}}$ & $7.2\pm1.3^{\textrm{a}}$ & $6.5\pm0.4^{\textrm{a}}$  
& $2.2\pm0.5$ & $6.1\pm 0.6$ & $5.5\pm2.9\dag$ 
& $17.2\pm1.0^{\textrm{a}}$ & $3.5\pm0.5$  \\
NGC 7469  &  $25.8\pm1.6$  & $4.5\pm0.3$     &    $<1.1$      &    
$12.4\pm 3.2$    &    $5.6\pm 1.3$    &     $<0.7$       
&    $6.1\pm 1.7$    &     $<1.1$       &    $11.4 \pm 
2.3$    &     $<1.9$       \\
NGC 7674  &  $32.1\pm1.1$  &  $3.1\pm0.3$    &    $1.5\pm 0.2$    
&    $8.3\pm 0.6$    &    $2.3\pm 0.8$    &     $<3.0$       
&    $4.1\pm 0.8$    &     $<1.4$       &    $3.5\pm 1.0$    
&   $<0.7$           \\
NGC 7682  &  $31.8\pm0.8$  & $1.9\pm0.1$     &     $<0.8$       &    
$3.0\pm 1.4$    &    $2.0\pm1.0$    &     $<0.4$       &    
 $<1.5$       &     $<1.8$       &     $<1.4$       &    
 $<1.5$       \\
\enddata
\tablenotetext{*}{Out of the spectral range}
\tablenotetext{**}{Fall within a  telluric zone}
\tablenotetext{***}{Fall in a CO band absorption zone}
\tablenotetext{\S}{Sum  of narrow and broad components}
\tablenotetext{\star}{Severely affected by atmospheric absorption}
\tablerefs{
\noindent \phantom{}$^{\textrm{a}}$ \citet{b27}, \hspace{0.1cm}\noindent 
\phantom{} $^{\textrm{b}}$ \citet{b33},  \hspace{0.1cm}\noindent 
\phantom{} $^{\textrm{c}}$ \citet{b50},  \hspace{0.1cm}\noindent 
\phantom{} $^{\textrm{d}}$ \citet{b71},   
\noindent \phantom{}$^{\textrm{e}}$ \citet{b100},  
\hspace{0.1cm}\noindent \phantom{} $^{\textrm{f}}$ \citet{b51}, 
\hspace{0.1cm}\noindent \phantom{} $^{\textrm{g}}$ \citet{b48}, 
\hspace{0.1cm}\noindent \phantom{} $^{\textrm{h}}$ \citet{b31}, 
\hspace{0.1cm}\noindent \phantom{} $^{\textrm{i}}$ \citet{b101}, 
\hspace{0.1cm}\noindent \phantom{} $^{\textrm{j}}$ \citet{b23}, 
\hspace{0.1cm}\noindent \phantom{} $^{\textrm{k}}$ \citet{b29},  
\hspace{0.1cm}\noindent \phantom{} $^{\textrm{l}}$ \citet{b53}      
}
\end{deluxetable}

\begin{deluxetable}{lccc|lccc} 
\tablewidth{0pt}
\tablecaption{Wavelengths, ionization Potential and critical densities  for the studied CLs*. Values for [S\,{\sc iii}]   and [Fe\,{\sc  ii}]   are included for purposes of comparison. \label{tab2}}   
\tablehead{
\colhead{Line}  &  \colhead{$\lambda$}     &  \colhead{$\chi$, IP}  &  \colhead{$\log n_{e}$}      &   \colhead{Line}   & \colhead{$\lambda$}   &  \colhead{$\chi$, IP}   &   \colhead{$\log n_{e}$}     \\  
\colhead{}  &  \colhead{($\mu$m)}  &  \colhead{(eV)} &  \colhead{(cm$^{-3}$)}  &   \colhead{}           & \colhead{($\mu$m)}  &  \colhead{(eV)}   &  \colhead{(cm$^{-3}$)}}
\startdata
[S\,{\sc  iii}]   &  0.9530  &  23.3    &  5.8  &  [Si\,{\sc  x}] &  1.4301 &   351.1 & 8.8 \cr
[S\,{\sc  viii}]  &  0.9913  &  280.9   &  10.6 &  [S\,{\sc  xi}] &  1.9196 &  447.1 & 8.5   \cr
[Fe\,{\sc  xiii}] &  1.0747  &  330.8   &  8.8  &  [Si\,{\sc  vi}] & 1.9630 &   166.8 & 8.8 \cr 
[S\,{\sc  ix}]    &  1.2520  &  328.2   &  9.4  &  [Al\,{\sc  ix}] & 2.0450 &   284.6 & 8.3 \cr 
[Fe\,{\sc  ii}]   &  1.2570  &  7.9     &  5.0  & [Ca\,{\sc  viii}] & 2.3213 &  127.7 & 7.9 \cr
\enddata
\tablenotetext{*}{Other most important CLs in the optical and NIR  referred to in this work:
[Si\,{\sc vii}] 2.48$\mu$m, IP=205 eV;
[Fe\,{\sc vii}] 6087~\AA\, IP=100 eV;
[Fe\,{\sc x}] 6374~\AA\, IP=235 eV;
[Fe\,{\sc xi}] 7889~\AA\ IP=262 eV}
\end{deluxetable}

\clearpage

\begin{deluxetable}{lccccccccccc}     
\tabletypesize{\scriptsize}
\rotate
\tablecaption{FWHM  (in km\,s$^{-1}$) of  coronal and low-ionization lines in the spectral window $0.8-$2.4~$\mu$m sorted by increasing value of IP. All data are from the sample of RRAP06 except when indicated. All values of FWHMs are instrument corrected ($\sim$ 360 km\,s$^{-1}$) except when  taken from the literature. The last column (T) indicates the general observed trend  between FWHM and the ionization potential (three or more data were required): I  (increase trend), P (increase trend up to 300 eV and then a decrease or constant),  L (level-off), N (no trend).  \label{tablex2}}       
\tablewidth{0pt}
\tablehead{
      &  \colhead{[Fe   II]}  & \colhead{[S III]} &  \colhead{[Ca VIII]} &   \colhead{[Si VI]} &   \colhead{[S VIII]}  &   \colhead{[Al IX]}  &  \colhead{[S IX]} & \colhead{[Fe XIII]} &  \colhead{[Si X]} & \colhead{[S XI]} &  T  \\  
     Galaxy          & \colhead{1.257 $\mu$m} &  \colhead{0.953  $\mu$m} &   \colhead{2.3213 $\mu$m}  &  \colhead{1.9630 $\mu$m}  &  \colhead{0.9913  $\mu$m}   &       \colhead{2.0450 $\mu$m}   &  \colhead{1.2520 $\mu$m}  &        \colhead{1.0747 $\mu$m}  &      \colhead{1.4301 $\mu$m}  &     \colhead{1.9196 $\mu$m} & }   
\startdata
Mrk 335                     &      $\cdots$           &      570$^{\textrm{a}}$  &    $\cdots$     &      830$^{\textrm{a}}$ &    1050$^{\textrm{a}}$    &    $\cdots$    &   \dag    &  1400$^{\textrm{a}}$&  1150$^{\textrm{a}}$ &     $\cdots$   &   P     \\
NGC 262                   &      $590\pm10$   &      $550\pm10$   &    $\dag$    &     $450\pm20$           &  $740\pm50$    &    $\cdots$    &    $530\pm10$       &        $\cdots$          &  $540\pm40$          &     $\cdots$    &  L  \\
Mrk 993                     &    $530\pm30$      &      $850\pm40$       &    $\cdots$     &    $\dag$     &  $\dag$    &    $\cdots$    &   $\cdots$     &     $\cdots$             &    $\cdots$  &   $\cdots$  &        \\
NGC 591                   &   $590\pm30$      &       $\leq360$              &    $\dag$      &   $400\pm20$    & $940\pm100$        &     $\cdots$   &   $\cdots$  &       $\cdots$          &  $\leq360$          &   $\dag$               & N   \\
Mrk 573                      &   $\leq360$           &        $\leq360$             &    $\leq360$   &   $\leq360$             &  $\leq360$       &   $\cdots$    & $\leq360$   &     $\cdots$           &   $\leq360$         &    $\leq360$                &    \\
NGC 863                    &   350$^{\textrm{a}}$     &        420$^{\textrm{a}}$              &       $\cdots$           &  350$^{\textrm{a}}$     &  600$^{\textrm{a}}$  &      $\cdots$        &  350$^{\textrm{a}}$       &    $\cdots$    &   \dag    &      $\cdots$   & P   \\
NGC 1068                  &  1070$^{\textrm{e}}$   &  $\ddag$   & 1060$^{\textrm{f}}$  &  950$^{\textrm{b}}$   & $\ddag$ &   $\cdots$         &  $\ddag$  & 1300$^{\textrm{e}}$ &   $\ddag$   &      $\cdots$     &  N  \\
Mrk 1066                    &   $500\pm20$  &    $530\pm10$              &     $\dag$     &   $460\pm60$            &     $\cdots$   &     $\cdots$     &   $\cdots$    &      $\cdots$ &     $\cdots$  &     $\cdots$  & L \\
MCG-05-13-017        &   $\leq360$    &        $\leq360$     &      $\dag$           &    $680\pm90$       &  $580\pm10$   &       $\cdots$        &   $\leq360$       &     $\cdots$              &    $450\pm120$         &      $\cdots$        & N   \\
ESO 428-G014         &   $540\pm40$ &        $420\pm10$               &    \dag           &   $610\pm20$       &   $840\pm100$       &     $\cdots$          &      $\cdots$           &     $\cdots$    &     $550\pm 70$    &    $\cdots$ &  N    \\
Mrk 78                         &        $\cdots$      &   $\cdots$      &   $\cdots$      &   $\ddag$     &    $\ddag$          &      $\cdots$    &    $\cdots$    &     $\cdots$  &    $\cdots$       &  $\cdots$      &                         \\
Mrk 1210                    &    $\cdots$   &     $780\pm10$         &    $\leq360$   &    $570\pm50$           &   $800\pm90$      &     $\cdots$         &    $860\pm70$     &      $\cdots$        &    $560\pm130$         &       $\cdots$       &  N  \\
Mrk 124                       & $990\pm160$  &        $550\pm30$             &       $\cdots$              &       $\dag$              &         $\dag$        &    $\cdots$  & $\cdots$   &   $\cdots$   &  $\cdots$   &  $\cdots$  &  \\
Mrk 1239                     &   $\leq360$      &        $580\pm160$            &    $770\pm30$             &    $1360\pm100$         &  $1600\pm80$    &       $\cdots$         &   $1790\pm140$      &      $\cdots$         &  $1520\pm80$        &       $\cdots$   &    P     \\ 
NGC 3227                   &   $700\pm40$   &          $660\pm30$             &     $\dag$          &     $610\pm110$          &  \dag    &    $\cdots$  &    $\cdots$  &      $\cdots$    &   $\dag$    &   $\cdots$  &  L  \\
PG 1126-041              &    $\cdots$   &       $550\pm20$             &     \dag \dag          &    $610\pm10$            &   $\dag$      &    $\cdots$         &     $\dag$           &       $\cdots$              &      $2130\pm170$        &     $\cdots$    &   I     \\
H1143-182                  &   $790\pm80$      &        $440\pm30$             &       $\cdots$           &   $740\pm40$             &      $\cdots$        &     $\cdots$   &   $400\pm20$    &     $\cdots$       &  $700\pm40$          &      $\cdots$   &   N \\
NGC 4051                    &   $\leq360$     &       $380\pm10$                &  \dag \dag \dag  &     $410\pm40$          &   $1150\pm90$      &   $\leq360$     &   $570\pm50$         &    $1110\pm140$      &  $460\pm10$    &   $\leq360$   & P   \\
NGC 4151                    &   $790\pm40$     &       $440\pm20$                &    $\leq360$              &     $\leq360$           &  $950\pm160$      &     $\dag$         &    $500\pm80$        &     $\cdots$    & $480\pm30$          &     $\dag$       &  N  \\ 
Mrk 766                        &   $\leq360$    &       $\leq360$                 &    $\leq360$            &      $\leq360$          &   $580\pm90$     &     $\cdots$    &   $760\pm70$         &    $\cdots$     & $470\pm50$         &       $\cdots$   &  P    \\
NGC 4748                    &   $\leq360$     &       $420\pm10$                &   $\dag$ &      $630\pm30$           &  $1740\pm300$   &     $\cdots$    &     $\cdots$       &     $\cdots$  &   $\cdots$   &      $\cdots$        &    I   \\ 
Mrk  279                        &  $580\pm30$     &       $850\pm30$                &    $\cdots$   &      $\dag$           &    $\dag$        &       $\cdots$        &  $\dag$       &     $\cdots$                &    $\cdots$         &         $\cdots$        &        \\
Circinus                        &  143$^{\textrm{d}}$ &      $\ddag$  & 150-540$^{\textrm{*h}}$  & 175-300$^{\textrm{*h}}$  &  $\leq100^{\textrm{g}}$  &     200-300$^{\textrm{*h}}$ &     97$^{\textrm{d}}$      &  $\cdots$   &    $\cdots$   &   $\cdots$    &  P   \\
NGC 5548                    &  $900\pm150$      &     $430\pm10$                 &    $500\pm20$             &     $940\pm110$           &   $990\pm100$       &      $\cdots$         &   $600\pm40$   &      $\cdots$     &  $470\pm30$     &  $\dag$   &   N    \\
NGC 5728                   &    $370\pm70$    &     $380\pm50$             &    $\cdots$      &   $510\pm40$            &    $\cdots$      &   $\cdots$     &    $\cdots$     &     $\cdots$            &    $\cdots$         &    $\cdots$            &     I   \\
PG 1448+273             &   $\cdots$     &     $530\pm40$               &    \dag \dag            &    $440\pm30$            & $\dag$       &     $\cdots$    &    $\cdots$     &     $\cdots$              &     $\cdots$        &       $\cdots$         &       \\
PG 1612+261             &   \dag \dag \dag  &      $490\pm10$   &   \dag \dag          &   $620\pm80$             &  $\dag$       &     $\cdots$     &   $\dag$  &      $\cdots$          &   $\dag$       &      $\cdots$  &         \\    
Mrk 504                        &    $\cdots$     &    $400\pm30$            &       $\cdots$          &      $\dag$          &   $\dag$       &      $\cdots$       &      $\cdots$        &     $\dag$         &     $\dag$       &     $\cdots$        &          \\
1H 1934-063A            & 550$^{\textrm{a}}$   &  420$^{\textrm{a}}$      &   350$^{\textrm{a}}$   &   650$^{\textrm{a}}$   &   400$^{\textrm{a}}$   &   $\cdots$   &   $580^{\textrm{a}}$  &  1600$^{\textrm{a}}$ & 1100$^{\textrm{a}}$ &  $\cdots$   &  P  \\
Mrk 509                        &        $\cdots$       &        $\cdots$                  &        $\cdots$                   &          500$\pm30$      &         $\cdots$          &  $\cdots$   &        $\dag$     &                $\cdots$              &          $\S$      &     $\cdots$  &    \\
Mrk 896                        &        $\cdots$     &     $\leq 360$        &     $\cdots$        &     $\dag$        &     $\cdots$        & $\cdots$            &    $\cdots$         &    $\cdots$         &    $\cdots$         &       $\cdots$               &   \\
1H 2107-097              &   $\cdots$    &    $\leq360$            &      $\dag$          &  $\leq360$             &  $\dag$     &  $\cdots$       &    $\dag$        &         $\cdots$              &    $\dag$      &       $\cdots$     &     \\
Ark 564                        &   350$^{\textrm{a}}$  &   440$^{\textrm{a}}$   &   350$^{\textrm{a}}$  &  530$^{\textrm{a}}$  &   890$^{\textrm{a}}$  &  $550\pm20$   &  480$^{\textrm{a}}$&920$^{\textrm{a}}$ &  590$^{\textrm{a}}$  &  $500\pm20$   &  P  \\
NGC 7469                   &  $\leq360$     &    $360\pm10$             &      \dag \dag \dag          & $\leq360$              & $840\pm50$        &     $\cdots$           & $580\pm90$           &      $\cdots$               &   $730\pm20$         &         $\cdots$    &    P          \\
NGC 7674                   &  $660\pm20$      &  $680\pm10$             &   $\leq360$              &  $550\pm110$             & $850\pm110$        &       $\cdots$          &   $1150\pm60$       &      $\cdots$             &  $920\pm160$          &      $\cdots$       &  N    \\
NGC 7682                                &   $380\pm40$    &      $\leq360$          &      $\cdots$         & $\leq360$              &  $\dag$      &      $\cdots$        &    $\cdots$        &       $\cdots$            &     $\cdots$      &        $\cdots$      &     L   \\ 
\enddata
\tablenotetext{*}{Decomposition  in two components}
\tablenotetext{\dag}{Strongly blended or difficult to measure}
\tablenotetext{\dag \dag}{Out of the spectral range}
\tablenotetext{\dag \dag \dag}{Severely affected by intrinsic or atmospheric absorption}
\tablenotetext{\ddag}{FWHM is not reported although the line exists}
\tablenotetext{\S}{Fall within a telluric zone}
\tablerefs{
\noindent \phantom{} $^{\textrm{a}}$ \citet{b27}, \hspace{0.1cm}\noindent \phantom{} $^{\textrm{b}}$ \citet{b50},  \hspace{0.1cm}\noindent \phantom{} $^{\textrm{c}}$ \citet{b33},  \hspace{0.1cm}\noindent \phantom{} $^{\textrm{d}}$ \citet{b98},    \\
\noindent \phantom{} $^{\textrm{e}}$ \citet{b99},  \hspace{0.1cm}\noindent \phantom{} $^{\textrm{f}}$ \citet{b29}, \hspace{0.1cm} \noindent \phantom{} $^{\textrm{g}}$ \citet{b23}, \hspace{0.1cm} \noindent \phantom{} $^{\textrm{h}}$ \citet{b101}
}
\end{deluxetable}



\begin{thebibliography}{}


\bibitem[Alexander, Young \& Hough(1999)]{b99} Alexander D. M., Young S.,  Hough J. H.   1999, \mnras, 304, L1
\bibitem[Appenzeller \& \"Ostreicher(1988)]{b67} Appenzeller I.,  \"Ostreicher R. 1988, \aj,  95, 45    
\bibitem[Bassani et al.(1999)]{b84} Bassani L.,  Dadina M.,  Maiolino R., Salvati M., Risaliti G., Della Ceca R., Matt G.,  Zamorani G. 1999, \apjs, 121, 473
\bibitem[Bianchi, Guainazzi \& Chiaberge(2006)]{bgc06} Bianchi, S., Guainazzi, M., Chiaberge, M. 2006, A\&A, 448, 499
\bibitem[Cappi et al.(2006)]{b77} Cappi M. et al. 2006,  A\&A, 446, 459
\bibitem[Cohen(1983)]{b65} Cohen R. D. 1983, \apj,  273, 489
\bibitem[Cooke et al.(1976)]{b14} Cooke  B. A., Elvis M., Maccacaro, T., Ward M. F., Fosbury A. E.,  Penston M. V.   1976, \mnras, 177, 121P
\bibitem[Crenshaw et al.(1991)]{b96} Crenshaw D. M., Peterson B. M., Korista K. T., Wagner R. M., Aufdenberg, J. P. 1991,  \aj,  101, 1202
\bibitem[De Robertis \& Osterbrock(1986)]{dro86} De Robertis, M. M., Osterbrock, D. E. 1986. \apj, 301, 727
\bibitem[Erkens, Appenzeller \& Wagner(1997)]{b25} Erkens U., Appenzeller I., Wagner  S.  1997, A\&A,  323, 707 
\bibitem[Evans(1988)]{b70} Evans I. N. 1988, \apjs, 67, 373

\bibitem[Ferguson et al.(1997)]{fer97} Ferguson, J. W., Korista, K. T., Ferland, G. J. 1997,
\apjs, 110, 287
\bibitem[Ferland(1993)]{b91} Ferland G. J.   1993, Department of Physics and Astronomy Internal Report, University of Kentucky
\bibitem[Geballe et al.(2009)]{g09} Geballe, T., Mason, R., Rodr\'{\i}guez-Ardila, A., Axon, D. 2010, \apj, 701, 1710
\bibitem[Gelbord et al.(2009)]{gel09} Gelbord, J. M., Mullaney, J. R., Ward, M. J. 2009, MNRAS, 397, 172
\bibitem[Giannuzzo, Rieke \& Rieke(1995)]{b24} Giannuzzo E., Rieke G. H.,  Rieke M. J.  1995, \apj, 446, L5
\bibitem[Glikman, Helfand \& White(2006)]{gli06} Glikman, E., Helfand, D. J., White, R. L. 2006, \apj, 640, 579

\bibitem[Grandi(1978)]{b12} Grandi S. A.  1978, \apj, 221, 501
\bibitem[Guainazzi, Matt \& Perola(2005)]{b83} Guainazzi M., Matt G.,  Perola G. C.  2005,  A\&A, 444, 119
\bibitem[Kaspi et al.(2005)]{b80} Kaspi S., Maoz D., Netzer H., Peterson B., Vestergaard M., Jannuzi B. T.  2005, \apj, 629, 61
\bibitem[Knop et al.(1996)]{b37} Knop R. A., Armus L., Larkin J. E., Matthews D. L., Shupe D. L., Soifer B. T.  1996, \aj, 112, 81
\bibitem[Korista \& Ferland(1989)]{b69} Korista K. T.,  Ferland G. J.  1989, \apj, 343, 678
\bibitem[Leighly(1999)]{b86} Leighly K. M. 1999, \apjs, 125, 317
\bibitem[Levenson, Weaver \& Heckman(2001)]{b82} Levenson N. A., Weaver K. A.,  Heckman T. M.  2001, \apj, 550, 230
\bibitem[Lutz et al.(2002)]{b26} Lutz D.,  Maiolino R., Moorwood A. F. M., Netzer  H., Wagner  S. J., Sturm  E., Genzel  R.  2002, A\&A, 396, 439 
\bibitem[Maiolino et al.(1998)]{b53} Maiolino R., Krabbe A., Thatte N., Genzel R.   1998, \apj, 493, 650
\bibitem[Marconi et al.(1996)]{b33} Marconi A.,  van der Werf P. P., Moorwood A. F. M.,  Oliva E.   1996, A\&A, 315, 335
\bibitem[Mazzalay \& Rodr\'{\i}guez-Ardila(2007)]{b48} Mazzalay X., \& Rodr\'{\i}guez-Ardila A.    2007, A\&A, 463, 445
\bibitem[Mazzalay, Rodr\'{\i}guez-Ardila \& Komossa(2010)]{m10} Mazzalay X., Rodr\'{\i}guez-Ardila A., Komossa, S. 2010, \mnras, 405, 1315 
\bibitem[Moorwood  et al.(1996)]{b93} Moorwood A. F. M., Lutz D., Oliva E., Marconi A.,  Netzer  H.,  Genzel R., Sturm E., de Graauw Th. 1996, A\&A, 315, L109
\bibitem[M\"ueller S\'anchez et al.(2006)]{b101} M\"uller-S\'anchez F., Davies R. I., Eisenhauer F., Tacconi L. J., Genzel R., Sternberg A.   2006,  A\&A,  454, 481
\bibitem[M\"ueller S\'anchez et al.(2009)]{mul09} M\"uller-S\'anchez, F., Davies, R. I., Genzel, R., Tacconi, L. J., Eisenhauer, F., Hicks, E. K. S., Friedrich, S., Sternberg, A. 2009, \apj, 696, 448
\bibitem[M\"ueller S\'anchez et al.(2011)]{mul11} M\"uller-Sánchez, F., Prieto, A., Hicks, E. K. S., Vives-Arias, H., Davies, R. I., Malkan, M., Tacconi, L. J., Genzel, R. 2011, Accepted in \apj, arXiv:1107.3140.
\bibitem[Murayama \& Taniguchi(1998)]{b41} Murayama T.,  Taniguchi Y.  1998,  \apj,  497, L9
\bibitem[Murayama,  Taniguchi \& Iwasawa(1998)]{b68} Murayama T.,  Taniguchi Y.,  Iwasawa K.  1998, \aj, 115, 460
\bibitem[Nagao, Taniguchi \& Murayama(2000)]{b40} Nagao T., Taniguchi Y.,  Murayama T. 2000, \aj, 119, 2605
\bibitem[Nandra \& Pounds(1994)]{np94} Nandra, K., Pounds, K. A. 1994, \mnras, 268, 405
\bibitem[Neumayer et al.(2007)]{neu07} Neumayer, N., Cappellari, M., Reunanen, J., Rix, H.-W., van der Werf, P. P., de Zeeuw, P. T., Davies, R. I. 2007, \apj, 671, 1329
\bibitem[Nucita et al.(2010)]{nuc10} Nucita, A. A., Guainazzi, M., Longinotti, A. L., et al. 2010, \aap, 515A, 47
\bibitem[Oke \& Sargent(1968)]{b11} Oke  J. B., Sargent W. L.  1968, \apj, 151, 807
\bibitem[Oliva et al.(1994)]{b23} Oliva E., Salvati M., Moorwood  A. F. M., Marconi A. 1994, A\&A, 288, 457
\bibitem[Oliva, Marconi \& Moorwood(1999)]{b5} Oliva E., Marconi A., Moorwood  A. F. M. 1999, A\&A, 342, 87
\bibitem[Oliva et al.(2001)]{b71} Oliva E.  et al.  2001,  A\&A, 369, L5 
\bibitem[Osterbrock \& Parker(1964)]{b6} Osterbrock D. E.,  Parker R. A. 1964, \apj, 141, 892
\bibitem[Osterbrock(1969)]{b2} Osterbrock D. E. 1969, Astrophys. Lett., 4, 57
\bibitem[Osterbrock(1981)]{b36} Osterbrock D. E. 1981,  \apj,  246, 696
\bibitem[Osterbrock \& Martel(1993)]{b94} Osterbrock D. E., Martel A. 1993,  \apj, 414, 552
\bibitem[Panessa et al.(2006)]{b79} Panessa F., Bassani L., Cappi M., Dadina M., Barcons X., Carrera F. J., Ho L. C., Iwasawa K.   2006, A\&A, 455, 173
\bibitem[Pelat, Alloin \& Fosbury(1981)]{b16} Pelat D.,  Alloin D.,   Fosbury R. A. E. 1981, \mnras, 195, 787
\bibitem[Pelat, Alloin \& Bica(1987)]{b66} Pelat D.,  Alloin D.,  Bica E.  1987, A\&A, 182, 9
\bibitem[Penston et al.(1984)]{b4} Penston M. V., Fosbury A. E., Boksenberg A., Ward M. J., Wilson A. S. 1984, \mnras, 208, 347
\bibitem[Pfefferkorn, Boller \& Rafanelli(2001)]{b57}  Pfefferkorn F.,  Boller Th.,  Rafanelli P.  2001, A\&A, 368, 797
\bibitem[Pier \& Voit(1995)]{b18} Pier E. A.,  Voit G. M.  1995, \apj, 450, 628
\bibitem[Pogge \& Owen(1993)]{b39} Pogge  R. W.,  Owen J. M.  1993, Ohio State Uni. Int. Rep., 93-01
\bibitem[Porquet et al.(1999)]{b60} Porquet D., Dumont A.M., Collin S., Mouchet M.  1999, A\&A, 341, 58
\bibitem[Prieto \& Viegas(2000)]{b28} Prieto M. A., Viegas S. M.  2000, \apj, 532, 238
\bibitem[Prieto, P\'erez Garc\'{\i}a \&  Rodr\'{\i}guez Espinosa(2002)]{b61} Prieto M. A, P\'erez Garc\'{\i}a A. M.,  Rodr\'{\i}guez Espinosa J. M.  2002, \mnras, 329, 309
\bibitem[Prieto et al.(2004)]{pri04} Prieto M. A., Meisenheimer, K., Marco, O., et al. 2004, \apj, 614, 135 
\bibitem[Prieto, Marco \&  Gallimore(2005)]{b52} Prieto M. A, Marco O., Gallimore J.  2005, \mnras, 364, L28
\bibitem[Ramos-Almeida et al.(2006)]{b51}  Ramos-Almeida C., P\'erez Garc\'{\i}a A.M., Acosta-Pulido J.A., Rodr\'{\i}guez Espinosa J.M., Barrena R., Manchado A. 2006, \apj,  645, 148
\bibitem[Reunanen, Kotilainen \& Prieto(2002)]{b34} Reunanen J., Kotilainen J. K., Prieto  M. A.  2002, \mnras, 331, 154
\bibitem[Reunanen, Kotilainen \& Prieto(2003)]{b29} Reunanen J., Kotilainen J. K., Prieto M. A.  2003, \mnras, 343, 192
\bibitem[Reynolds(1997)]{r97} Reynolds, C. S. 1997, \mnras, 286, 513
\bibitem[Riffel, Rodr\'{\i}guez-Ardila \& Pastoriza(2006)]{b30} Riffel R., Rodr\'{\i}guez-Ardila A., Pastoriza M. G.  2006, A\&A, 457, 61
\bibitem[Riffel et al.(2009)]{rif09} Riffel, R., Pastoriza, M. G., Rodr\'{\i}guez-Ardila, A., Bonatto, C. 2009, \mnras, 400, 273
\bibitem[Rodr\'{\i}guez-Ardila et al.(2002)]{b27} Rodr\'{\i}guez-Ardila A., Viegas S. M., Pastoriza  M., Prato L. 2002, \apj, 579, 214
\bibitem[Rodr\'{\i}guez-Ardila, Contini \& Viegas(2005)]{b31} Rodr\'{\i}guez-Ardila A., Contini M., Viegas S. M. 2005, \mnras, 357, 220
\bibitem[Rodr\'{\i}guez-Ardila et al.(2006)]{b50} Rodr\'{\i}guez-Ardila  A.,  Prieto M. A.,  Viegas S. M., Gruenwald R. 2006, \apj,  653, 1098
\bibitem[Rose et al.(2011)]{ros11} Rose, M., Tadhunter, C. N., Holt, J., Ramos Almeida, C., Littlefair, S. P. 2011, \mnras, 414, 3360
\bibitem[Rush et al.(1996)]{b58} Rush B., Malkan M. A., Fink H. H., Voges W. 1996, \apj, 471, 190
\bibitem[Sambruna et al.(2001a)]{s01a} Sambruna, R. M., Brandt, W. N., Chartas, G., Netzer, H., Kaspi, S., et al. 2001, \apj, 546, L9 
\bibitem[Sambruna et al.(2001b)]{s01b} Sambruna, R. M., Netzer, H., Kaspi, S., Brandt, W. N., Chartas, G.,  et al. 2001, \apj, 546, L13 
\bibitem[Shields \& Oke(1975)]{b3} Shields G. A., Oke  J. B. 1975, \apj, 1975 
\bibitem[Shinozaki et al.(2006)]{b81} Shinozaki K., Miyaji T., Ishisaki Y., Ueda Y., Ogasaka Y. 2006, \aj, 131, 2843
\bibitem[Schlesinger et al.(2009)]{schl09} Schlesinger, K., Pogge, R. W., Martini, P., 
Shields, J. C., Fields, D. 2009, \apj, 699, 857
\bibitem[Shuder(1980)]{b95} Shuder J. M. 1980, \apj, 240, 32 
\bibitem[Shuder \& Osterbrock(1981)]{b64} Shuder J. M., Osterbrock  D. E. 1981, \apj, 250, 55 
\bibitem[Storchi-Bergmann et al.(1999)]{b98} Storchi-Bergmann T., Winge C., Ward M. J., Wilson A. S. 1999, \mnras, 304, 35
\bibitem[Terashima et al.(2002)]{ter02}Terashima, Yuichi; Iyomoto, Naoko; Ho, Luis C.; Ptak, Andrew F.
\bibitem[Thompson(1995)]{b35} Thompson R. I.  1995, \apj, 445, 700 
\bibitem[Thompson(1996)]{b100} Thompson R. I.  1996, \apj, 459, L61 
\bibitem[Viegas-Aldrovandi \& Contini(1989)]{b44} Viegas-Aldrovandi S. M.,  Contini M. 1989, A\&A, 215, 253
\bibitem[Walter \& Fink(1993)]{b59} Walter R.,  Fink H. H. 1993, A\&A, 274, 105
\bibitem[Ward \& Morris(1984)]{b17} Ward M.,  Morris S. 1984, \mnras, 207, 867
\bibitem[Wilson(1979)]{b15}  Wilson A. S.  1979,   Proc. R.  Soc.  Lond., A366, 461
\bibitem[Whittle et al.(2005)]{whi05} Whittle, M., Rosario, D. J., Silverman, J. D., 
Nelson, C. H., Wilson, A. S. 2005, \aj, 129, 104


\end{thebibliography}
\end{document}